\documentclass[fleqn,usenatbib]{mnras}

\usepackage{newtxtext,newtxmath}

\usepackage[T1]{fontenc}
\usepackage{float}

\DeclareRobustCommand{\VAN}[3]{#2}
\let\VANthebibliography\thebibliography
\def\thebibliography{\DeclareRobustCommand{\VAN}[3]{##3}\VANthebibliography}

\usepackage{graphicx}	\usepackage{subfig}
\usepackage{amsmath}	\usepackage{color}
\usepackage{tablefootnote}

\newcommand{\refone}[1]{{#1}}

\title[ExSeSS AGN number counts]{The Extragalactic Serendipitous Swift Survey (ExSeSS) -- I. Survey definition and measurements of the X-ray number counts}

\author[Jack N. Delaney et al.]{
Jack N. Delaney$^{1}$,\thanks{E-mail: delaney@roe.ac.uk}
James Aird$^{1,2}$,
Phil A. Evans$^{2}$,
Cassandra Barlow-Hall$^{1}$,
Julian P. Osborne$^{2}$
and \newauthor
Michael G. Watson$^{2}$
\\
$^{1}$Institute for Astronomy, University of Edinburgh, Royal Observatory, Blackford Hill, Edinburgh, EH9 3HJ\\
$^{2}$University of Leicester, University Rd, Leicester, LE1 7RH\\
}

\date{Accepted 2022 November 23. Received 2022 November 23; in original form 2022 July 28.}
\pubyear{2022}

\begin{document}
\label{firstpage}
\pagerange{\pageref{firstpage}--\pageref{lastpage}}
\maketitle

\begin{abstract}
We present the Extragalactic Serendipitous Swift Survey (ExSeSS), providing a new well-defined sample constructed from the observations performed using the Swift X-ray Telescope. The ExSeSS sample consists of 79,342 sources detected in the medium (1-2~keV), hard (2-10~keV) or total (0.3-10~keV) energy bands, covering 2086.6~deg$^{2}$ of sky across a flux range of  $f_\mathrm{0.3-10keV}\sim10^{-15}-10^{-10}$ erg~s$^{-1}$~cm$^{-2}$. Using the new ExSeSS sample we present measurements of the differential number counts of X-ray sources as a function of 2-10~keV flux that trace the population of Active Galactic Nuclei (AGN) in a previously unexplored regime.
We find that taking the line-of-sight absorption column density into account has an effect on the differential number count measurements and is vital to obtain agreement with previous results.
In the hard band, we obtain a good agreement between the ExSeSS measurements and previous, higher energy data from \textit{NuSTAR} and \textit{Swift}/BAT when taking into account the varying column density of the ExSeSS sample as well as the X-ray spectral parameters of each of the samples we are comparing to. We also find discrepancies between the ExSeSS measurements and AGN population synthesis models, indicating a change in the properties of the AGN population over this flux range that is not fully described by current models at these energies, hinting at a larger, moderately obscured population at low redshifts ($z\lesssim0.2$) that the models are not currently taking into account.

\end{abstract}


\section{Introduction}

Active Galactic Nuclei (AGN) occur when the super-massive black hole (SMBH) at the centre of a galaxy is rapidly accreting dust and gas, emitting large amounts of energy across the electromagnetic spectrum. Identifying AGN at X-ray wavelengths is particularly effective as the X-ray emission is able to penetrate surrounding material, allowing obscured sources that would not be found at other wavelengths to be identified \citep{brandt2015cosmic, netzer2015revisiting, hickox2018obscured}. Observing in the X-ray band also enables the identification of low-luminosity AGN that would be too faint and diluted by the host galaxy light at other wavelengths. Finally, the vast majority of detected point sources in X-ray surveys are associated with AGN. All these factors make X-ray surveys ideal for identifying SMBHs in their growth phase as an AGN.

Since X-rays provide an extremely efficient way of probing the AGN population, significant efforts have been dedicated to a variety of different surveys. \textit{ROSAT} \citep{trumper1982rosat} performed the first all-sky survey over the 0.1--2~keV energy band  with the sensitivity of a focusing X-ray telescope, identifying over 100,000 sources \citep{voges2000rosat,boller2016second}.
The \textit{eROSITA} instrument on the Specktrum Roentgen-Gamma mission launched in 2019 has carried out a new, high-sensitivity all-sky X-ray survey at 0.2-8~keV, repeatedly scanning the sky every 6 months and is expected to detect $>$3 million AGN, with the majority identified in the most sensitive, soft energy band \citep[0.2--2.3~keV,][]{merloni2012erosita,predehl2021erosita}.

While \textit{ROSAT} and \textit{eROSITA} probe the whole sky, \textit{Chandra} and \textit{XMM-Newton} enable much deeper surveys over smaller areas. In particular, \textit{Chandra} has carried out surveys ranging from $\sim$200~ks to 7~Ms depth over fields of $\sim$0.1--0.6~deg$^2$  \citep[][]{alexander_Chandra_2003,laird_AEGISX_2009,xue_Chandra_2011, nandra_AEGISX_2015,luo_Chandra_2017, kocevski_XUDS_2018}.
The deepest \textit{Chandra} survey, the Chandra Deep Field South \citep[CDF-S:][]{luo2016chandra} covers 484.2 arcmins$^{2}$ for 7Ms of exposure with a total of 1008 sources detected across multiple energy bands, reaching an X-ray point source density of $\sim$ 23,900 deg$^{-2}$ for AGN.
These deep fields are complemented by wider area, shallower surveys including the $\sim$2~deg$^2$ COSMOS-Legacy survey \citep[reaching $\sim$160ks depth per pointing and containing $\sim$4000 sources:][]{civano2016chandra} and the 9.3 deg$^{2}$ Chandra Deep Wide Field Survey \citep[CDWF-S, reaching $\sim$30~ks depth per pointing with a total of 6891 sources:][]{masini2020chandra}. \textit{XMM-Newton} has a larger collecting area and a wider field-of-view, but lacks the angular resolution of \textit{Chandra}, so instead efforts have focused on performing shallower surveys over larger areas of sky \citep[e.g. the $\sim$50~deg$^2$ XMM-XXL survey:][]{2016A&A...592A...1P}.

In addition to these dedicated survey efforts, extremely large X-ray samples can also be constructed using
sources found in the field-of-view during observations of dedicated targets with a different science objective.
The large field-of-view of \textit{XMM-Newton} makes it especially suitable for constructing such samples (e.g.~2XMM:\citealt{watson2009xmm}; 3XMM:\citealt{rosen2016xmm}) with the latest compilation
\citep[4XMM:][]{webb2020xmm}
containing 550,124 X-ray sources and covering a total sky area of $\sim$1152~deg$^2$. A similar effort, the \textit{Chandra} Source Catalog \citep[CSC:][]{chen2019vizier,evans_2010} contains a total of 315,000 source covering a total area of $\sim550$ deg$^{2}$ in its latest release \footnote{https://cxc.cfa.harvard.edu/csc/char.html}.
However, 4XMM and CSC contain both sources associated with the targets of the observation and serendipitously detected sources within the field-of-view.
We note that a key challenge in the analysis of these X-ray catalogues is the identification and removal of any X-ray sources that are associated with the target to construct a truly serendipitous sample. \citet{mateos2008high} describes the process of making an extragalactic serendipitous sample using 2XMM, producing a catalogue of 1129 \textit{XMM-Newton} sources from a total sky area of 132.3~deg$^2$.

While sensitive surveys at soft X-ray energies ($\lesssim2$~keV) predominantly identify unobscured AGN, surveys at harder energies ($\sim$2--10~keV) are able to identify both low and moderately obscured sources, although remain biased against the most heavily obscured (Compton-thick) sources. Surveys at even higher X-ray energies ($>$10keV) are less biased against such populations.
The Burst Alert Telescope (BAT) on the \textit{Neil Gehrels Swift Observatory} \citep[hereafter \textit{Swift},][]{barthelmy2005burst} is sensitive to an energy range of 15--150~keV and is constantly observing a large fraction of the sky to identify gamma ray bursts (GRBs) and measure their positions on the sky to $\sim$4-arcminute accuracy. \citet{oh2018105} present the most recent (105 month) catalogue containing 1632 persistent hard X-ray sources identified in the 14--195~keV energy band. However, \textit{Swift}/BAT only detects the brightest sources in the local universe due to its comparatively poor angular resolution and limited sensitivity.
The \textit{Nuclear Spectroscopic Telescope Array} (\textit{NuSTAR}) X-ray Observatory, launched in 2012, was the first mission with grazing-incidence mirrors capable of focusing high-energy ($\sim$10--80~keV) X-rays \citep[][]{harrison2013nuclear,madsen2015calibration}, enabling surveys that detect much fainter sources at these energies, albeit over substantially smaller areas of sky \citep[$\sim$0.3--1.7~deg$^2$, see][]{mullaney15nustar,civano2015nustar,masini2018nustar}. To access larger areas of sky, \citet{lansbury2017nustar} constructed a catalogue of 497 truly serendipitous sources (i.e.~excluding targets) in the 3--24~keV, 3--8~keV or 8--24~keV bands using a compilation of the first 40 months of \textit{NuSTAR} observations, covering a total sky area of $\sim16$~deg$^2$ \citep[see also][]{alexander2013nustar}.

As X-ray point source samples in high Galactic latitude fields will be dominated by distant AGN, the most immediate quantification that an X-ray survey can provide---the number counts of sources in a given energy band at different fluxes---already places important constraints on the AGN population. To constrain the intrinsic X-ray source density as a function of flux (also referred to as the ``logN-logS'') requires a combination of deep, small-area surveys and wide, shallow surveys to probe a wide range of fluxes, as well as an accurate quantification of the sensitivity of the X-ray observations.  Measurements with \textit{Chandra} and \textit{XMM-Newton} have shown that the logN-logS is well described by a broken power-law over a broad range in flux, with a steep slope at brighter fluxes and a shallower slope at fluxes $\lesssim10^{-14}$~erg~s$^{-1}$~cm$^{-2}$ \citep[e.g.][]{mateos2008high,georgakakis2008new}. With additional data---most crucially redshifts of the X-ray sources---our observational picture can be expanded to include the luminosity function and its evolution \citep[e.g.][]{ueda2014toward,aird2015evolution}. By incorporating constraints on the fraction of sources with different levels of absorption, a full ``population synthesis model'' of AGN can be constructed that describes the evolution of AGN over cosmic time \citep[e.g.][]{gilli2007synthesis, ballantyne2014average} and can be tested using new measurements of the number counts in different energy bands.
\citet{harrison2016nustar} presented the first measurements of the X-ray source number counts based on the \textit{NuSTAR} surveys, including both the dedicated survey fields and the serendipitous sample from \citet{lansbury2017nustar}.
They found that the \textit{NuSTAR} 8--24~keV logN-logS agrees well at fluxes $\sim10^{-14}-10^{-12}$ erg~s$^{-1}$~cm$^{-2}$ with the predictions of population synthesis models that are based on earlier studies with \textit{Chandra} and \textit{XMM-Newton}.
However, the \textit{NuSTAR} number counts exceed the simplest extrapolation of the \textit{Swift/BAT} number counts at brighter fluxes and---most notably---the \textit{Swift}/BAT number counts do not agree with the population synthesis models that are successful at fainter fluxes, suggesting evolution of the AGN population between the higher redshifts sampled by \textit{NuSTAR} and the local universe sampled by \textit{Swift}/BAT that is not fully captured in these models.
However there have been few X-ray surveys that cover large enough area to directly probe the flux range spanning between the \textit{NuSTAR} and \textit{Swift}/BAT surveys, either at the same high energies ($\gtrsim10$~keV) or more moderate energies ($\sim$2--10~keV) to help resolve this discrepancy.

In this paper, we present the Extra-galactic Serendipitous Swift Survey (ExSeSS), a new sample of truly serendipitous sources at high Galactic latitudes and covering an area of $2086.6$~deg$^2$ which is constructed from observations carried out with the \textit{Swift} X-ray telescope (XRT).
Our sample is extracted from the second \textit{Swift} X-ray Point Source catalogue \citep[2SXPS:][]{evans20202sxps} which contains all sources detected in the 0.3--1~keV, 1--2~keV, 2--10~keV and 0.3--10~keV range, which we reduce to be primarily extragalactic (by excluding Galactic latitudes $-20^\circ<b<20^\circ$) and serendipitous to provide an unbiased sample of X-ray sources to study the AGN population.
In Section \ref{sec: Defining ExSeSS} we define the ExSeSS sample, describing the process to select the appropriate fields, identify and remove sources associated with the targets of the observations, create the source sample, and determine the overall area coverage and sensitivity.
The ExSeSS source catalogue is described in Appendix \ref{apend:source_catalogue} and made available online at {\url{TBD}}.
In Section \ref{sec: Number counts} we use the ExSeSS sample to measure the logN-logS with a particular focus on the 2--10~keV band where we probe a key flux range between \textit{NuSTAR} and \textit{Swift}/BAT, albeit at lower energies. We also compare ExSeSS to the established population synthesis models from  \citet{gilli2007synthesis}, \citet{ueda2014toward} and the updated model from
\citet{ballantyne2014average} as presented in \citet{harrison2016nustar}. Our results provide new insights to help understand the obscuration properties of the AGN over this important flux range. In Section \ref{sec: Conclusion} we provide a summary of our work and conclusions, as well as discussing the usefulness of the ExSeSS sample for future studies.

\section{Swift data and definition of the ExSeSS sample}\label{sec: Defining ExSeSS}

The sources included in this paper are taken from the 2nd Swift X-ray Point Source catalogue \citep[2SXPS:][]{evans20202sxps} which we have reduced to construct the Extragalactic Serendipitous Swift Survey (ExSeSS) sample. This process involved taking the full 3790 deg$^{2}$ covered by 2SXPS, removing areas covered by the Galactic plane or large nearby galaxies so that the sample is dominated by background extragalactic sources, and removing any X-ray sources associated with the targets from the resulting catalogues so that only truly serendipitous detections are retained.  We describe the process to construct the ExSeSS sample and our method to define the area coverage (to different flux limits) covered by our new survey.

\subsection{Swift/XRT Data}

The 2SXPS data consist of a list of datasets, which correspond to the individual observations carried out by \textit{Swift}/XRT, and a list of detections, which contain the thousands of sources that are detected in the different observations and in different energy bands (thus, in some cases, containing multiple detections of the same source).
Each dataset covers a region with a diameter of $\sim 23.6$', corresponding to the \textit{Swift}/XRT field-of-view.
Additional datasets were also generated by stacking the individual observations where overlapping areas of the sky have been viewed multiple times.
This stacking improves the sensitivity in that area, revealing fainter sources and increasing the number of sources in the catalogue.
In total in 2SXPS there are $143,697$ datasets containing $1,091,058$ independent detections in either the 0.3--10~keV (total), 0.3-1~keV (soft), 1--2~keV (medium) or 2--10~keV (hard) energy bands, corresponding to 206,335 distinct sources and reaching down to flux limits of $f_\mathrm{0.3-10keV}\sim10^{-14}$~erg~cm$^{-2}$~s$^{-1}$ \citep[][]{evans20202sxps}.

\subsection{Construction of the ExSeSS Sample}\label{Sec: defining exsess}

\subsubsection{Defining the ExSeSS fields}\label{subsec: Defining fields}

Defining ExSeSS involves selecting the fields (and corresponding 2SXPS datasets) to include. We started by removing all the datasets that go in to a stacked dataset, i.e.~the individual datasets that have been observed over roughly the same area of sky that were combined to form a stacked dataset. We retained the stacked datasets, corresponding to the deepest available data for a given area of sky, ensuring that we do not duplicate areas where multiple datasets cover the same location and that we have a well-defined sensitivity in these areas. Once this is done we are left with stacked datasets and the remaining unstacked datasets that are never used in a stack. Keeping both of these we have 18,640 datasets left in our sample.

\begin{figure*}
	\includegraphics[width=\textwidth]{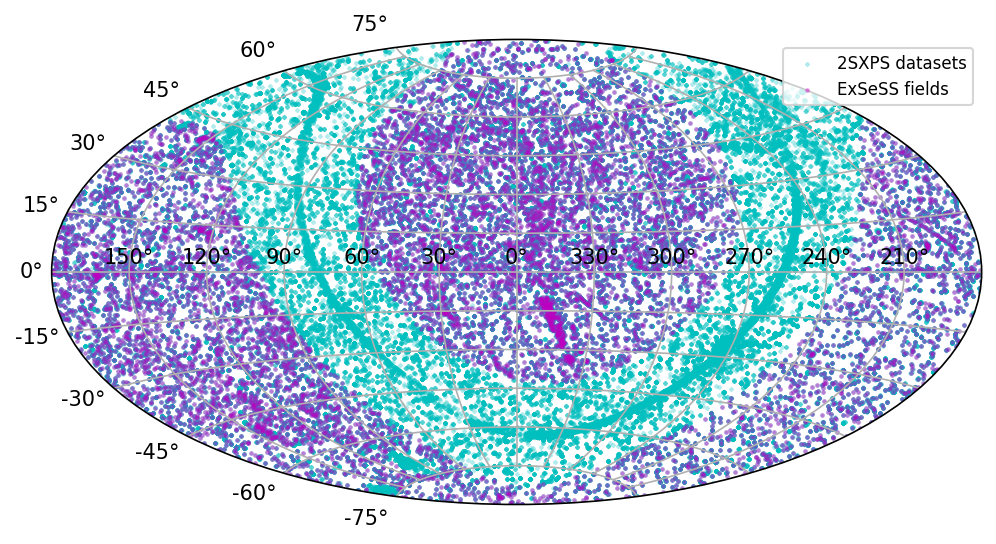}
    \caption{A map showing the distribution of the ExSeSS fields across the sky in right ascension and declination compared to the original 2SXPS. The 2SXPS datasets are shown in cyan and are distributed evenly over most of the sky. The ExSeSS fields are highlighted in purple and exclude the Galactic plane as well as areas of the sky such as the SMC and LMC. The process to define the ExSeSS fields is described in Section \ref{subsec: Defining fields}. }
    \label{fig:Sky_Map}
\end{figure*}

Any datasets within 2SXPS that contain diffuse emission were removed, as well as datasets with significant stray light or containing very bright sources with fitting issues, which is indicated by a  ``Field Flag'' not equal to zero in 2SXPS, leaving us with 17,514 datasets.
We also removed any datasets that lie within the Galactic plane (Galactic latitudes $-20^\circ<b<20^\circ$) to ensure our sample is dominated by extragalactic X-ray sources. Next, we removed any datasets that fall within the sky area of a number of well-known nearby objects: the Small and Large Magellanic Clouds (SMC and LMC); M31 and M33. These were selected as they are relatively close to our galaxy and have well known sizes. These objects are likely to contain many individually resolved X-ray sources such as stars and X-ray binaries (XRBs) which we do not want in our final sample. Our final list consists of $11,047$ datasets that are retained in ExSeSS and hereafter referred to as the ExSeSS ``fields'' (i.e. the areas of sky that form the ExSeSS survey). Table~\ref{tab:fields} summarises the definition of the ExSeSS fields, while Figure \ref{fig:Sky_Map} shows the distribution of these fields across the sky compared to the original 2SXPS datasets.

\begin{table}
\begin{center}
\begin{tabular}{ |p{3cm}||p{2.5cm}|}
 \hline
 \multicolumn{2}{|c|}{ExSeSS fields} \\
 \hline
 Fields type   & {Number of datasets \mbox{remaining}}\\
 \hline
 2SXPS original & 143,697 \\
 Unique Sky pointings  & 18,640  \\
 Good Field Flag   &  17,514  \\
 Large Objects removed  &  11,047  \\

 \hline
\end{tabular} \
 \caption{ Number of 2SXPS datasets that remain after applying each of the steps described in Section~\ref{subsec: Defining fields}. Originally there were 143,697 datasets in the 2SXPS sample. We remove individual datasets that are incorporated into the stacked fields, leaving us with 18,640 unique sky pointings. After applying the ``Field Flag'' to only use fields with good quality data we are left with 17,514 datsets. When removing the large objects (Galactic plane, SMC, LMC, M31, M33) we are left with 11,047 datasets, hereafter referred to as the ``ExSeSS fields''. In Section \ref{subsec: Defining Targets} if a target is larger than \textit{Swifts} field of view then the dataset will also be removed.
}\label{tab:fields}
\end{center}
\end{table}

\subsubsection{Identifying Swift targets for exclusion from the ExSeSS sample}\label{subsec: Defining Targets}

To create a truly serendipitous sample of sources requires that we identify the targets of the \textit{Swift}/XRT observations corresponding to each of our datasets and subsequently remove any X-ray sources associated with these targets.
The majority of our targets were identified using the database maintained by Pennsylvania State University\footnote{\url{https://www.swift.psu.edu/too_api/}}, which provides a list of the known targets of pointed observations.
We also removed GRB afterglows \footnote{\url{https://www.swift.ac.uk/xrt_positions}} that were identified using \textit{Swift}/XRT following triggered observations. The number of targets identified from  \refone{this process} are given in Table \ref{tab:Target_tables}.
Astronomical objects corresponding to these targets were then extracted from the SIMBAD\footnote{\url{http://simbad.u-strasbg.fr/simbad/sim-fbasic}} database using a closest match to find the angular size of the targets on the sky.
We adopt the major axis of the object as provided by SIMBAD as an estimate of the target's radius.

We manually checked all targets with a radius larger than 10 arcminutes to make sure that our SIMBAD cross-matching was identifying the correct target and found that in the majority of cases that the target did indeed have a large angular size (e.g. the M15 globular cluster or the M101 pair of galaxies were identified as targets, which contain multiple X-ray sources within the target radius that should be completely removed from our serendipitous source sample).
However there were 14 targets that were erroneously associated with an object with an extremely large angular size that is unlikely to correspond to an X-ray source or \textit{Swift} target (e.g. the Fermi bubble or nebula gas clouds). For these sources the true target could easily be identified with SIMBAD and they were manually updated.

Finally, we checked fields where no targets had been identified in the Pennsylvania State database. Overall there were 7,526 datasets that did not have a target associated with them following our initial analysis. To tackle this, we first identified the datasets that have zero detections and checked their exposure times. We thus identified 4,886 relatively shallow datasets with no source detections.
These fields correspond to observations where \textit{Swift} was tracking an object, such as a comet, for a short amount of time across the sky.
These fields are still very useful as they build up our overall sky coverage and are therefore kept in ExSeSS.
After this we were still left with 2,640 datasets with detections but no associated targets.
For these remaining datasets we examined the name assigned to the observation by the original observers and were able to identify a target on this basis.
In many cases, the target name was written in a non-standard format, requiring manual intervention to identify the correct target.
Other issues included follow-up of sources from the \textit{Swift}/BAT catalogue where we updated to the counterpart position from \citet{swift_bat_105}\footnote{\url{https://swift.gsfc.nasa.gov/results/bs105mon/}}.
\refone{
We caution that in a number of cases the target does not fall at the nominal pointing position.
Based on these manual checks, we identified 1007 extra targets to remove from 2SXPS to ensure a serendipitous sample, including the 14 that were identified in the Penn State database but were initially associated with a counterpart with angular size $>10$ ~arcminutes.
We then followed the same process of identifying a SIMBAD counterpart to determine the angular size of the additional
targets and added these to our list of targets and sky areas to exclude from ExSeSS.}

\refone{Of the remaining 977 datasets that still lacked targets, twelve corresponded to dedicated survey fields such as COSMOS (thus all sources can be treated as serendipitous detections) and the rest were `cooling' or `offset' pointings (where \textit{Swift} is purposefully pointed at a blank field position) or there was no other clearly defined target for the observation.
These datasets are retained in ExSeSS but are not assigned a target. }
Thus, our final ExSeSS data consists of 11,047 fields (defined in Section~\ref{subsec: Defining fields} above) with 12,224 targets to be removed. We have more targets than datasets because of the stacked datasets, which are made up of a combination of pointings, each with their own individual targets, and therefore one stacked dataset can have multiple targets within it.

\begin{table} \label{tab:Target_tables}
\begin{center}
\begin{tabular}{|p{5cm}||p{2cm}|}
 \hline
 \multicolumn{2}{|c|}{Targets identified} \\
 \hline
 Target list   & No. of Targets\\
 \hline
 Penn State target database & 10,561  \\
 GRB afterglows   &  656   \\
 Manually identified based on target name & \refone{1,007} \\
 \hline
 Total no. of targets & \refone{12,224} \\
 \hline
 Empty datasets not assigned targets & 4,886 \\
 Datasets without a defined target & \refone{977} \\

\hline
\end{tabular}

    \caption{Number of targets of \textit{Swift}/XRT observations that are identified via different routes. The largest number of targets are identified from Penn State database.         Additional targets were identified from the list of known GRB afterglows that were the subject of \textit{Swift}/XRT follow-up. \refone{Finally, we manually identified an additional 1,007 targets based on the target name recorded by the original observers. In total there are 12,224 targets to be excluded from ExSeSS to make the sample serendipitous. There are an additional 5,863 datasets that lack a defined target or contain no source detections and we do not assign a target.}}
\label{tab:Target_tables}
\end{center}
\end{table}

\subsubsection{Defining the ExSeSS sources}\label{subsec: Defining Sources}

\begin{figure*}
	\includegraphics[width=\textwidth]{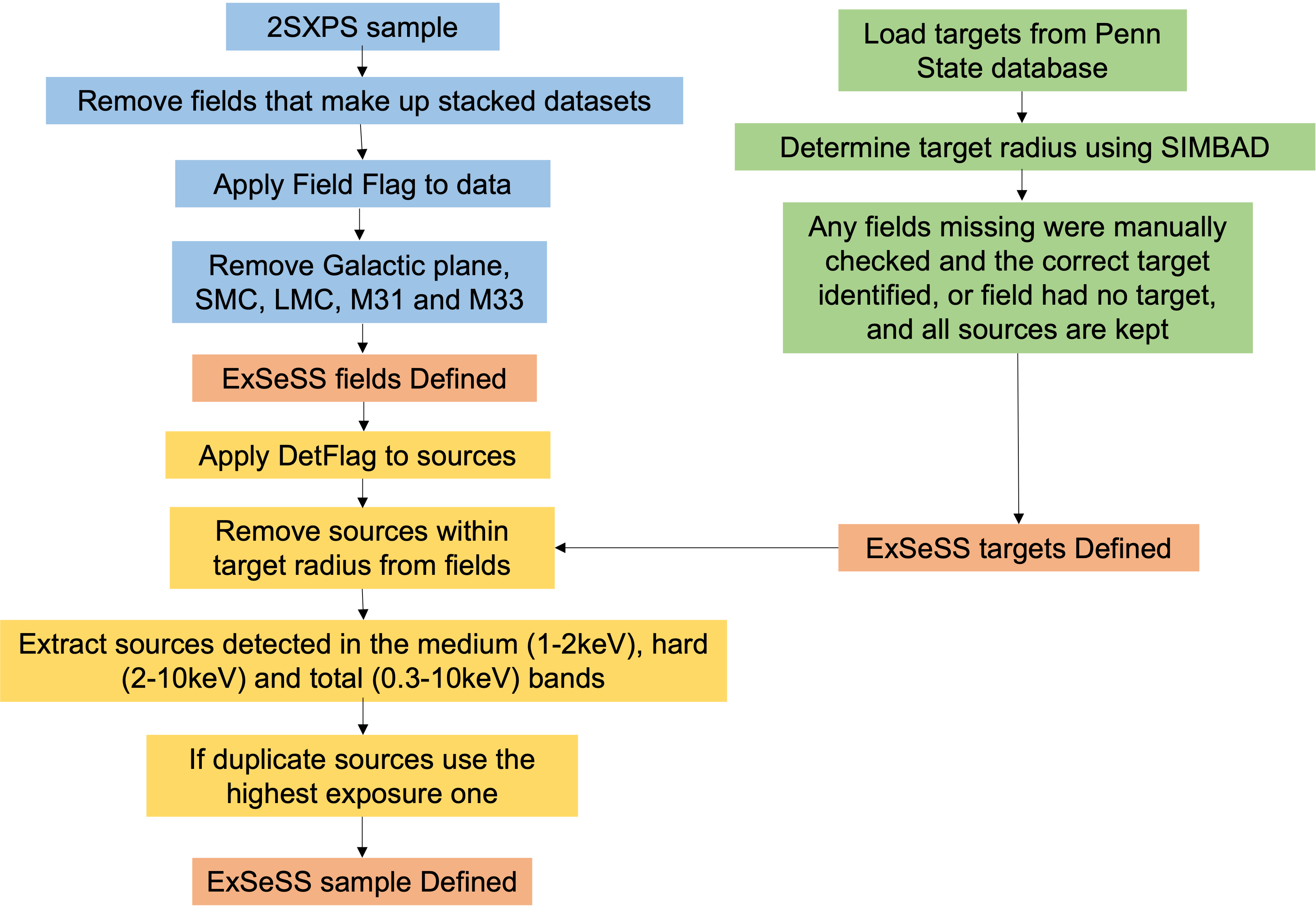}
    \caption{A flow diagram showing the process to create the ExSeSS sample from the 2SXPS data. In blue is the process used to define the ExSeSS fields and ensure the sample is dominated by distant, extragalactic sources. In green is the process to identify the targets of \textit{Swift}/XRT observations and determine their angular sizes. The yellow boxes show the process to define the ExSeSS source sample and ensure it contains only serendipitous detections. The orange boxes indicate the key outputs from the process: the list of fields, the list of targets, and the final ExSeSS source sample. The process is explained in more detail in Section \ref{Sec: defining exsess}.}
    \label{fig:flow diagram}
\end{figure*}

Our next step was to determine which X-ray detections from within the much larger catalogue provided by 2SXPS should be included in the ExSeSS sample of sources. First, we extracted only those 2SXPS detections that are associated with the refined list of ExSeSS fields, as described in Section~\ref{subsec: Defining fields} above, ensuring that our sample was dominated by extragalactic sources.
\refone{Indeed, we expect our sample to be dominated by distant AGN \citep[based on extrapolations of the X-ray number counts of different non-AGN populations from][see also Section \ref{sec: Number counts} below]{lehmer2012}, although detailed cross-matching and classification using multiwavelength data (deferred to a future work) is required to confirm the level of contamination.}
At this stage, we also limited our sample to ``good'' detections in the 2SXPS catalogue, which reduces the false detection rate to 0.3\% \citep[see][for more details]{evans20202sxps}. Where the edges of fields overlap some duplicate sources remain; in these cases we retained the detection with the highest exposure.

To ensure that our sample is serendipitous, we must remove any detected sources that are associated with the target of an observation, which would otherwise severely bias our sample.
We removed any X-ray detections that lie within a radius of a target object corresponding to the angular size (as defined in Section~\ref{subsec: Defining Targets} above) and thus ensured that sources that are associated with the target do not contaminate our sample. We also apply a conservative minimum radius of 2 arcminutes to prevent any contamination from detections of the target source itself or spurious detections in the wings of bright targets. The process for creating the ExSeSS sample is summarised in Figure \ref{fig:flow diagram} as a flow diagram to illustrate the process of taking the 2SXPS sample and refining it to construct the ExSeSS sample.

Finally, we limited our sample to sources detected in the medium (1-2~keV), hard (2-10~keV) or total (0.3-10~keV) bands from 2SXPS
leaving us with a total of 79,342 sources in the ExSeSS sample.
We converted the rates in the individual bands into fluxes assuming a simple power law X-ray spectrum with a photon index of $\Gamma = 1.9$ and correcting for Galactic absorption with $N_\mathrm{H}=4\times10^{20}$~cm$^{-2}$, which corresponds to the average over the ExSeSS area.
Fluxes were estimated in the standard 0.5--2~keV band based on the medium (1--2~keV) band count rates to enable comparison with prior X-ray surveys, as well as in the 2--10~keV and 0.3--10~keV energy bands\footnote{Conversion factors were calculated using the WebPIMMS tool: \url{https://heasarc.gsfc.nasa.gov/cgi-bin/Tools/w3pimms/w3pimms.pl}}.
This differs from the way 2SXPS calculated fluxes, which used spectral fits or hardness ratios of the individual sources. We do not use the same approach as 2SXPS here as assuming a single spectral shape means that a consistent count-to-flux conversion is applied both when measuring fluxes for individual sources and when accounting for the sensitivity, as well as avoiding effects due to poorly constrained spectral properties that come from individual hardness ratios in 2SXPS.
In Section \ref{sec: Variable_column} we use a more sophisticated approach to calculate the hard band fluxes, which involves adding an absorption factor calculated from the average hardness ratios.

\begin{figure}
	\includegraphics[width=\columnwidth]{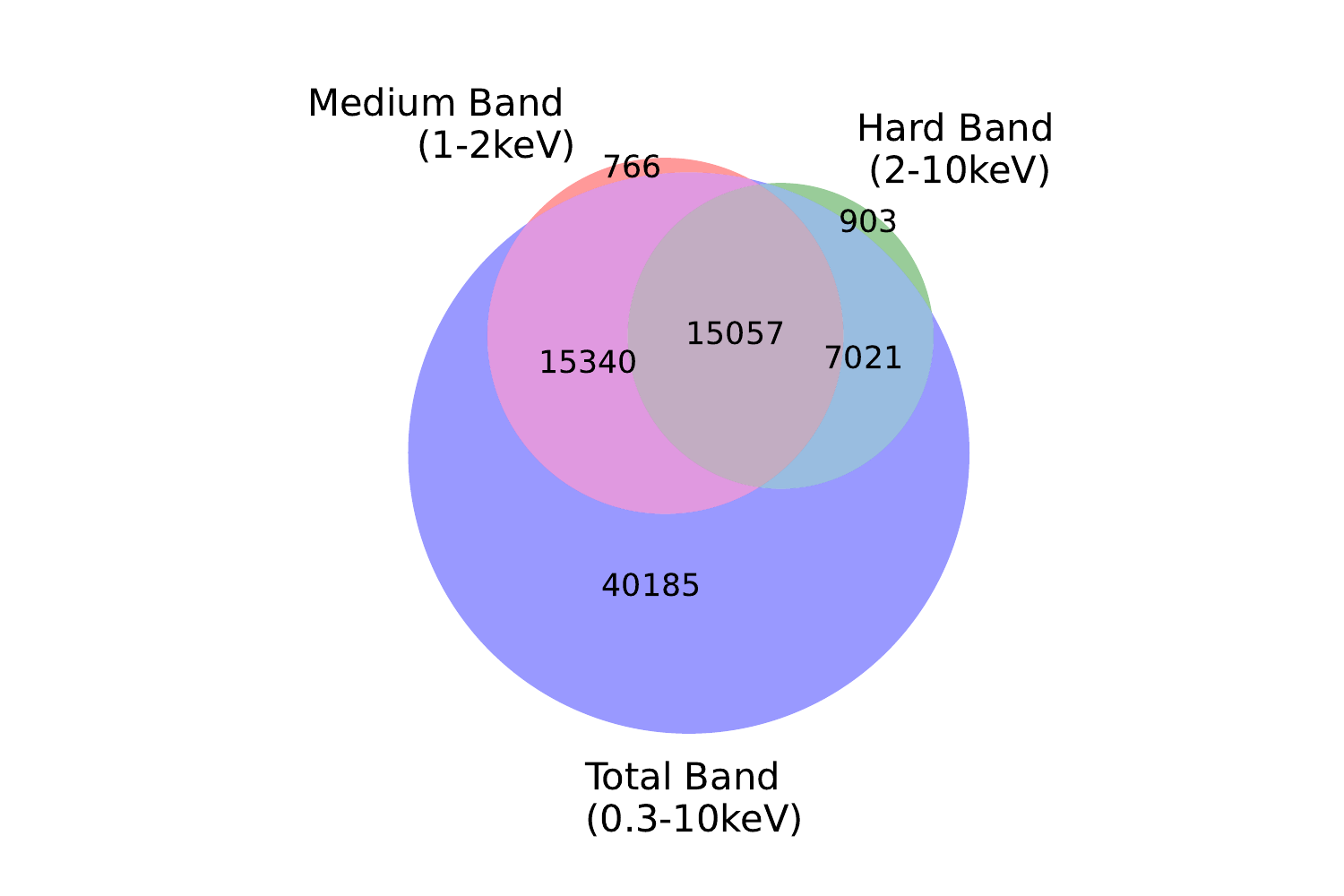}
    \caption{Venn diagram showing the number of sources that are detected in the medium (red), hard (green) and total (blue) bands, with overlapping regions indicating the numbers of sources detected in multiple bands.
        The total band has the largest number of sources as it has the best sensitivity but there are 766 sources detected only in the medium band and 903 sources detected only in the hard band.     }
    \label{fig:Venn diagram}
\end{figure}

Figure \ref{fig:Venn diagram} provides a Venn diagram showing the number of sources in the medium, hard and total band and the overlap between them when sources are detected in multiple bands.
For example, the brown overlapping region indicates the 15,057 sources that are detected in all three bands. Most of the ExSeSS sources are detected in the total band but there are a small number of sources that are only detected in the medium or hard bands.

\begin{figure}
	\includegraphics[width=\columnwidth]{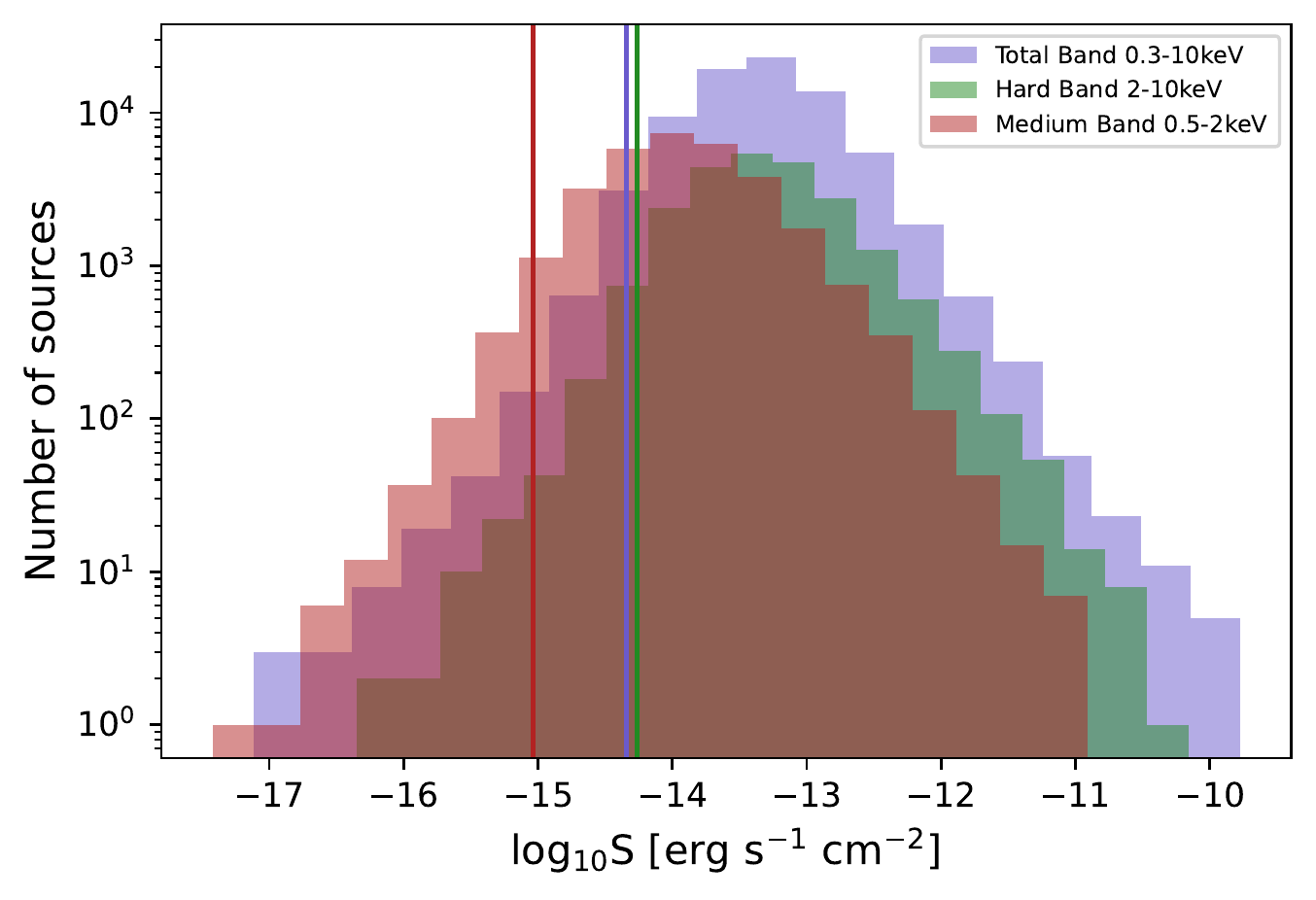}
    \caption{Distribution of fluxes for sources detected in each band. The total band (blue) has the largest number of sources covering a broad range of fluxes. The hard band (green) has more sources at brighter fluxes compared with the medium band (red) which has more sources at fainter fluxes. The solid lines show the reliable sensitivity limit at 0.1\% area cut-off; below this limit the area curve becomes unreliable and we thus exclude sources with fluxes fainter than this limit from our statistical analysis (see Section \ref{sec: Number counts}). }
    \label{fig:Histogram}
\end{figure}

 Figure \ref{fig:Histogram} shows how the number of sources are distributed in each band in terms of flux, with the largest number of sources detected in the total band.
 In the remainder of this paper we will mainly focus on the hard (2-10~keV) band sample as such a large sample over this flux range is unique to ExSeSS.
A description of the final source catalogue is given in Appendix~\ref{apend:source_catalogue}, while the full catalogue is made available online.\footnote{\url{https://www.swift.ac.uk/2SXPS/exsess/}}

\subsection{Calculation of survey area and sensitivity}
\label{sec: area curve}

\begin{figure*}
    \centering
	\includegraphics[width=\textwidth,trim=0.5cm 0 0cm 0cm]{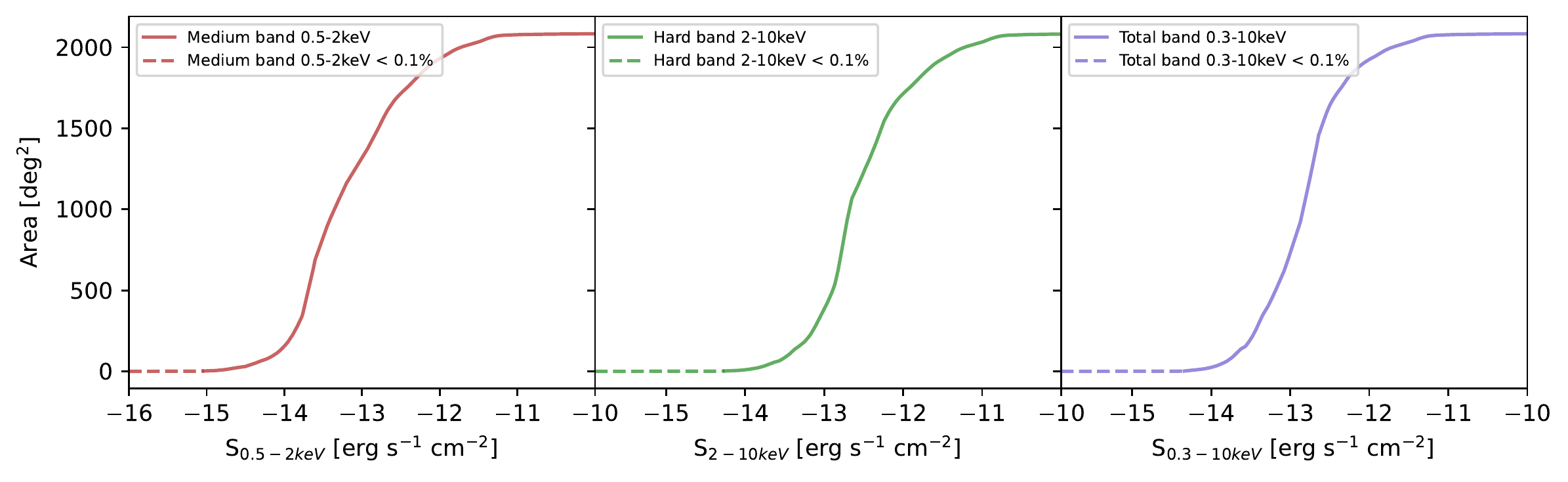}
    \caption{Area of sky covered by ExSeSS as a function of flux for the medium, hard and total band samples. Fluxes are converted from count rates with a fixed spectral assumption (a power law with $\Gamma = 1.9$ and Galactic absorption).
    In the hard band, the area curve is shifted toward higher fluxes (compared to the medium or total band) due to the reduced sensitivity of the \emph{Swift}/XRT at higher energies.
        We note that the medium band area curve has different shape as it is more strongly affected by the Poisson nature of the detection and the transition between photon limited and background limited regimes due to the narrower (1--2~keV) band used for the detection. The dashed lines indicate where the area drops to below 0.1\% of the maximum value of the area and we are no longer able to accurately determine the sensitivity from our simulations. }
    \label{fig:Area_curve}
\end{figure*}

A major advantage of our careful process to define the ExSeSS fields (see Section \ref{subsec: Defining fields}) is that we end up with a survey with a well-defined area coverage. Furthermore, ensuring we only retain detections in stacked datasets (where available) ensures a well-defined sensitivity over any part of the sky, allowing us to generate area curves that give the flux limit reached as a function of sky area, as shown in Figure \ref{fig:Area_curve}.
To determine these area curves, we first calculated the sky area coverage as a function of exposure time based on the exposure maps for each of the ExSeSS fields, excluding any areas within the specified radius of targets that are also excluded in the construction of the source sample.
Secondly, we calculated the fraction of sources that would be detected in areas with a given exposure time as a function of their count rate, based on the simulations described in section 7 of \citet{evans20202sxps}.
We note that these simulations were carried out assuming different source and background count rates and must be scaled to predict the detection probability in specific energy bands at a certain exposure time given knowledge of the spectral shape of the background from 2SXPS.
The final step involved combining the area as a function of exposure time and the detection probability for a given exposure time to calculate the area as a function of source count rate.
 This process is then repeated for each band.
Figure \ref{fig:Area_curve} shows the resulting area curves, applying our standard conversion of rate to flux (assuming a $\Gamma = 1.9$ and Galactic absorption of $N_\mathrm{H}=4\times10^{20}$~cm$^{-2}$). However, we note that the sensitivity is determined in terms of count rate which is thus independent (to first order) of the assumed X-ray spectrum.

The total area coverage is $2086.6$ deg$^{2}$ making ExSeSS one of the largest, sensitive, truly serendipitous X-ray surveys to-date, especially in the hard (2--10~keV) band.
For each band there is a well defined area curve, which is needed to determine the true sky density of sources as a function of flux.
The dashed line on Figure \ref{fig:Area_curve} refers to where the area curve in each band goes below 0.1\% of its maximum value. While sources are detected with lower fluxes, below this point the simulations from \citet[][]{evans20202sxps} are insufficient to accurately trace the shape of the sensitivity curve (requiring many more simulations on a much refined grid of exposure times that is computationally unfeasible) and thus we class the area curve as unreliable below this point and do not use sources below these limits in our statistical analysis (see Section~\ref{sec: Number counts} below).

\subsection{Comparison of survey area and sensitivity with previous surveys }

\begin{figure}
	\includegraphics[width=\columnwidth,trim=0 0 1.2cm 0]{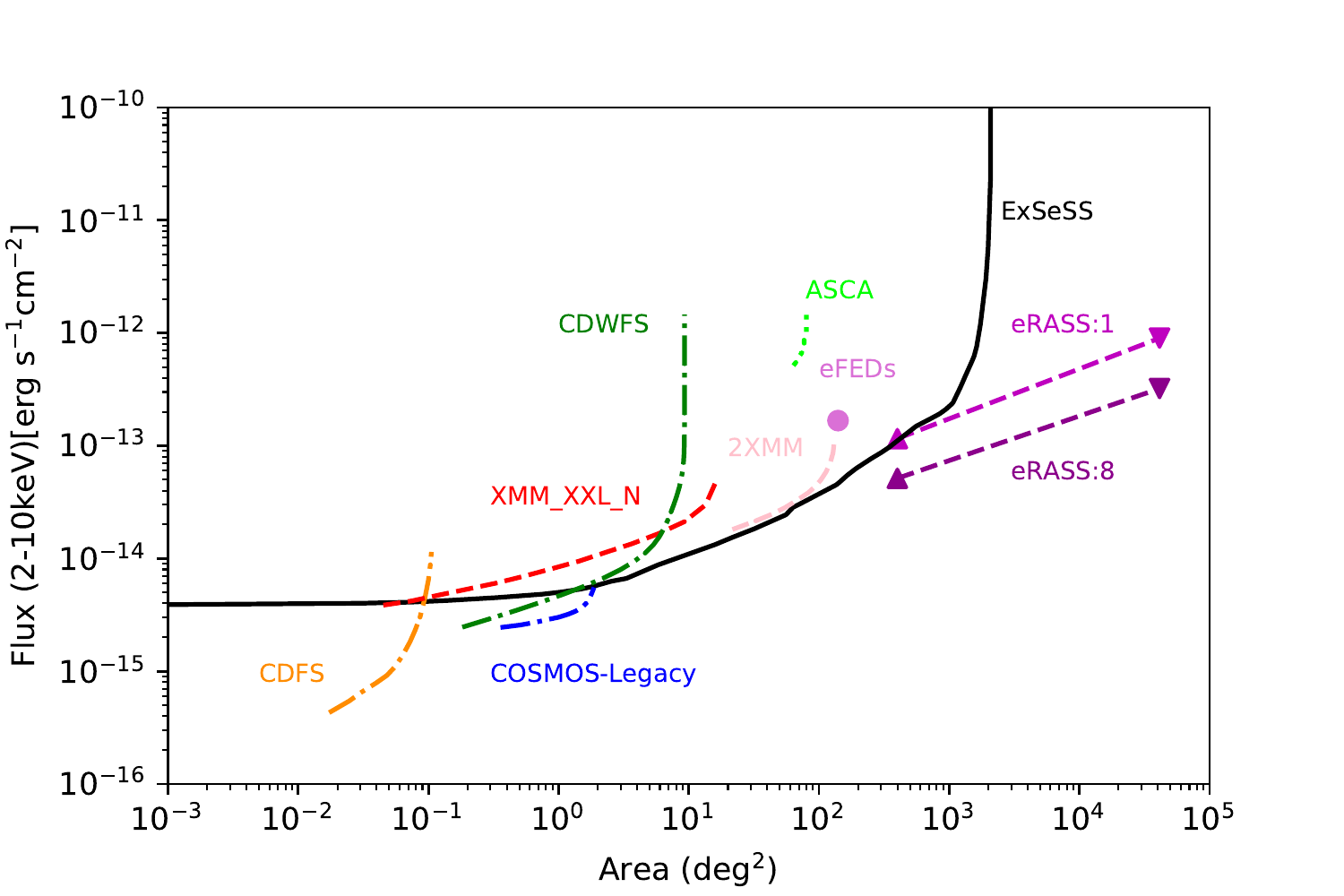}
    \caption{ExSeSS area coverage as a function of 2--10~keV flux limit (solid black line) compared with other surveys probing a similar energy flux range (curves span the flux limits achieved over 10\% to 90\% of the total area of a given survey).
    In red (XMM-XXL-N) and pink (2XMM) we have area coverage of two different \textit{XMM-Newton} surveys \citep[][]{liu2016x, mateos2008high}.
    Yellow (CDFS), blue (COSMOS-Legacy) and dark green (CDWFS) curves are from \textit{Chandra} surveys \citep[][]{luo2016chandra, civano2016chandra, masini2020chandra}.
    In lime green (ASCA) is the area coverage of the ASCA Large Sky Survey \citep[][]{ueda2001asca, ueda2005asca, akiyama2003optical, ueda2003cosmological}. ExSeSS has a very large area when compared to these other surveys and acts as a great precursor for \textit{eROSITA} \citep[][]{sunyaev2021srg}. The light purple point indicates the flux limit achieved in the 140 deg$^{2}$ eFEDS field \citep[][]{brunner2021erosita}. The purple line (eRASS:1) is sky coverage of eROSITA after it first scans the whole sky (with the triangle indicating the deeper limits for the equatorial poles) and the dark purple (eRASS:8) is the sky coverage at the final depth for the full 4-year survey. We note that \textit{eROSITA} surveys use the 2.3--5~keV band for source detection due to limited sensitivity above 5~keV, compared with ExSeSS which uses the full 2--10~keV band.}
    \label{fig:area_comparison}
\end{figure}

Figure \ref{fig:area_comparison} shows how ExSeSS compares to other surveys in terms of the sky coverage as a function of the hard (2--10~keV) band flux. It shows that ExSeSS has a much greater area coverage than prior surveys and covers a range of flux and area coverage that would otherwise remain unexplored for this energy band until \textit{eROSITA} finishes its all-sky survey, shown with the purple lines. Although \textit{eROISTA} will cover more sky to fainter fluxes, the sample is constructed in the 2.3--5~keV band due to limited sensitivity above 5~keV, whereas ExSeSS uses the data over the full 2--10~keV range and thus corresponds to a harder energy band. Thus, ExSeSS will still access a distinct parameter space even when the full-depth \textit{eROSITA} survey has been completed.

\section{Measurements of X-ray source number counts}\label{sec: Number counts}
\label{sec:results}

\subsection{Integrated number counts, $N(>S)$, in three energy bands}\label{sec: logN-logS}

One of the primary results that can be obtained from the carefully constructed ExSeSS sample is a measurement of the number count (sky density) of sources as a function of flux, known as the logN-logS.
To make these measurements, we first converted the count rates into fluxes for each band, initially assuming all sources have an X-ray spectrum with $\Gamma=1.9$ and Galactic absorption of $N_\mathrm{H}=4\times10^{20}$~cm$^{-2}$ only as described in Section \ref{subsec: Defining Sources}.

\begin{figure}
	\includegraphics[width=\columnwidth]{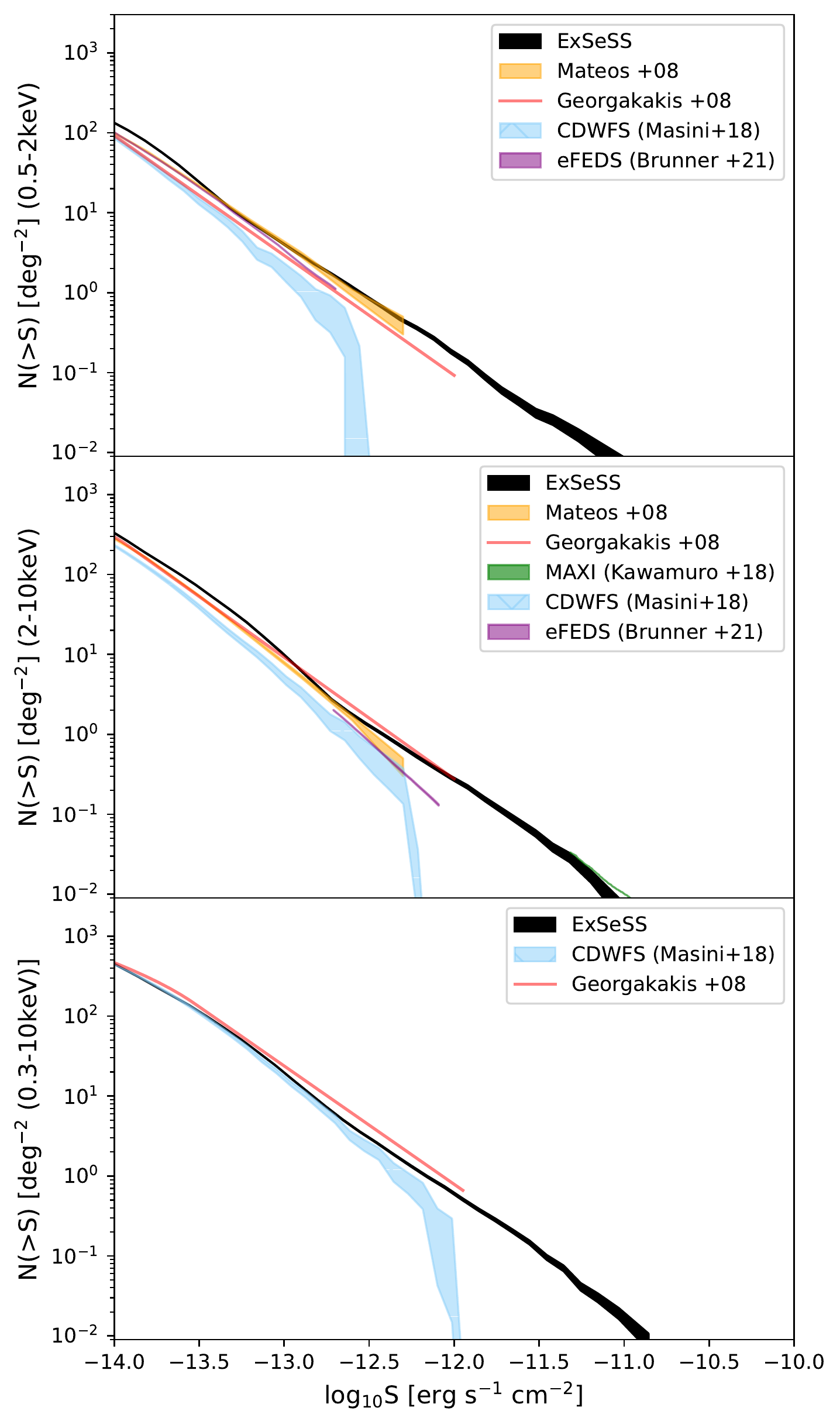}
    \caption{
    Integrated number counts, $N(>S)$, as a function of flux in the 0.5--2~keV (top), 2--10~keV (centre) and 0.3--10~keV (bottom) energy bands, based on the medium, hard and full band ExSeSS samples, respectively. We compare to previous measurements with \textit{XMM-Newton} \citep[][]{mateos2008high}, \textit{Chandra} \citep{georgakakis2008new}, \textit{eROSITA} in the $\sim140$~deg$^2$ eFEDS field \citep{brunner2021erosita}, \textit{Chandra} from the CDWFS \citep{masini2020chandra} and MAXI \citep{kawamuro20187}.
    For the medium band sample (comparing 0.5--2~keV fluxes) there is a good agreement between the results but the ExSeSS sample tends to give higher number counts below $10^{-13}$~erg~s$^{-1}$~cm$^{-2}$.
    In the hard band, ExSeSS agrees well with the previous results, including MAXI at very bright fluxes ($\sim10^{11}$erg~s$^{-1}$~cm$^{-2}$) and is higher than the CDWFS measurements at $\sim10^{-12.5}$ erg~s$^{-1}$~cm$^{-2}$.
    In the total band, ExSeSS is in good agreement with the CDWFS measurements \citep[][]{masini2020chandra} and lies slightly below the \citet{georgakakis2008new} best-fit relation at brighter fluxes.
    }
    \label{fig:logN-logS}
\end{figure}

The integrated number counts i.e. the number of sources greater than a given flux, $S$, is given by
\begin{equation}\label{equation: 1}
N(>S_{j}) =  \sum_{i=1}^{i=M} \frac{1}{\Omega_i}
\end{equation}

where the sum is taken over all sources with fluxes  $S_{i} > S_{j}$ and $S_{j}$ is the flux of the faintest object in the bin \citep[][]{mateos2008high}. $\Omega_i$ is the area coverage associated with source $i$ with flux $S_i$, obtained from Figure \ref{fig:Area_curve}, and allows us to account for the changing area that our survey is sensitive to throughout our flux range.
The error in the integrated number counts is given by $N(>S_{j})/M^{\frac{1}{2}}$ from Poisson statistics, where $M$ is the total number of sources with S$_{i}$ > S$_{j}$.
We have not included sources with fluxes that lie below the limit corresponding to 0.1\% of the total area as the area curve is likely to be unreliable below this level (see Section~\ref{sec: area curve} above).
Applying this cut removed $\sim$3\% of the sources in each band of our sample. This cut is applied to the sample for the rest of the paper.
Measuring the sky density of sources helps us to understand how the AGN population changes over different flux ranges. Figure \ref{fig:logN-logS} shows the logN-logS measurements for the ExSeSS sample and compares them to fits from \citet[][]{georgakakis2008new} and \citet{brunner2021erosita} and measurements from \citet[][]{masini2020chandra} and \citet{mateos2008high}.

The top plot of Figure \ref{fig:logN-logS} shows our measurements of $N(>S)$ as a function of 0.5--2~keV flux based on our medium band (1--2~keV) selected sample, compared to previous studies. We find that the ExSeSS sample is generally in good agreement with previous measurements, although with slightly higher number densities at both the high and low ends of our flux range.
We note that the measurements from the CDWFS \citep{masini2020chandra} give lower number densities at bright fluxes, which could be due to the relatively small size of the field.
The middle plot in Figure \ref{fig:logN-logS} shows the hard (2--10~keV) band ExSeSS number counts which also lie above the 2--10~keV measurements by \citet[][]{masini2020chandra} in the CDFWS and \citet[][]{brunner2021erosita} in the eFEDS field but are generally in good agreement with \citet[][]{georgakakis2008new} and \citet[][]{mateos2008high}.
 For the hard band, we also compare to results from MAXI \citep[][]{kawamuro20187} at the bright fluxes. There is a small dip in the ExSeSS number counts at these brighter fluxes and therefore our measurements lie slightly lower than the MAXI results but are reasonably consistent.
 We note the relatively small range of fluxes at these harder energies probed by the early \textit{eROSITA} final equatorial depth (eFEDS) performance verification data \citep{brunner2021erosita} and that a simple extrapolation of the best fitting power law found in that study would significantly under-predict the number counts at brighter fluxes compared to our ExSeSS measurements or the MAXI measurements.
 We investigate the hard band number counts in more detail, in their differential form, and compare to studies at higher energies in Section~\ref{sec: dN/dS} below.
 Finally the bottom plot in Figure \ref{fig:logN-logS} shows that there is some agreement between ExSeSS and previous measurements for the total 0.3--10~keV band with the \citet[][]{georgakakis2008new} fit slightly over-predicting the number density and the \citet[][]{masini2020chandra} measurements in good agreement with ExSeSS up to $10^{-12.5}$~erg~s$^{-1}$~cm$^{-2}$ where ExSeSS has greater area coverage and better source statistics.

\subsection{Average absorption column density as a function of flux for the hard-band ExSeSS sample}\label{sec: Variable_column}

To gain a greater understanding of the underlying AGN population, we now explore the spectral properties of the ExSeSS sample and use these constraints to get more accurate values for fluxes, rather than using a simple conversion that assumes a power law spectrum with $\Gamma = 1.9$. We focus on the hard band (2--10~keV) for the rest of this paper as such a large sample at these energies is a key feature of ExSeSS and substantially obscured AGN populations are more likely to be identified in this band.\footnote{
\refone{We choose not to calculate average HR and the equivalent effective $N_\mathrm{H}$ for the 3\% of hard-band sources in ExSeSS with fluxes below the 0.1\% area coverage limit as these faint sources tend to have poorly constrained HR values and are not included in our logN-logS measurements.}}
More careful consideration of spectral properties is also necessary when comparing in detail to previous
measurements both in this energy band and extrapolated from higher energies (see Section~\ref{sec:higherenergies} below).

\begin{figure}
	\includegraphics[width=\columnwidth]{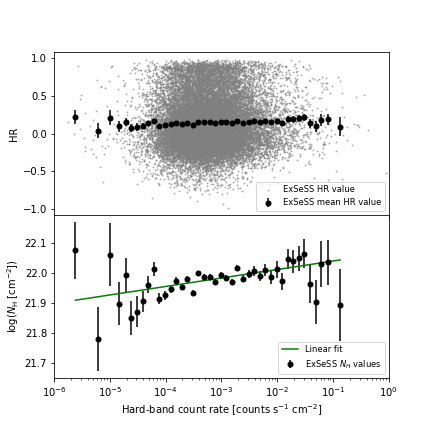}
    \caption{The first panel shows how the hardness ratio (HR) varies with rate across all of the ExSeSS sources (grey points) and the average in the pre-defined bins in hard-band count rate (black points). The second panel shows the $N_\mathrm{H}$ corresponding to the average HR as a function of count rate. The green line is a $\chi^2$ fit to the data with a linear relation in $\log N_\mathrm{H}--\log \mathrm{rate}$ space. On average the ExSeSS hard-band sample can be described by an X-ray spectrum with $\Gamma=1.9$ and an $N_\mathrm{H}$ $\approx$ 10$^{22}$~cm$^{-2}$. }
    \label{fig:Rate_vs_Nh}
\end{figure}

To take in to account how the spectral properties change over our hard-band sample, we used the hardness ratios (HRs) provided in the 2SXPS catalogue, which were calculated using
\begin{equation}
    HR = \frac{H - M}{H + M}
\end{equation}
where $H$ is the hard band (2--10~keV) count rate and $M$ is the medium band (1--2~keV) count rate.
Given the large scatter and uncertainty in individual HR values, we binned our sources according to the hard band count rate and calculated the mean HR for each bin. The top panel of Figure \ref{fig:Rate_vs_Nh} shows both the individual HR values for each source (grey crosses) and the mean HR values at a given hard band rate (black circles). The error in the mean HR values are given by the standard error in the mean.

We then used WebPIMMS\footnote{https://heasarc.gsfc.nasa.gov/cgi-bin/Tools/w3pimms/w3pimms.pl} to determine the HR that would be observed with \textit{Swift}/XRT for a range of $N_\mathrm{H}$ values and a fixed $\Gamma = 1.9$ spectrum.
We interpolated the above conversions to infer a mean $N_\mathrm{H}$ corresponding to the observed mean HR values at a given hard band count rate.
Figure \ref{fig:Rate_vs_Nh} (lower panel) shows how the inferred $N_\mathrm{H}$ changes as a function of count rate, where the green line is a $\chi^2$ fit to the data for a linear relation \refone{given by}
\begin{equation}
    \log \left(N_\mathrm{H} \;[\mathrm{cm^{-2}}]\right)= 0.0173 \log \left(\mathrm{count\; rate
    \; [s^{-1}]}\right) + 22.03.
\end{equation}
We used this linear fitting relation to assign a value of $N_\mathrm{H}$ to each of the individual hard-band ExSeSS sources, according to their count rate.
Taking an average of these values, we found that ExSeSS has a mean $N_\mathrm{H} = 10^{22}$~cm$^{-2}$.
Using the values of $N_\mathrm{H}$ assigned to each source, along with an assumed $\Gamma = 1.9$, provides flux conversion factors that take into account the impact of absorption on the \emph{typical} spectral properties of the ExSeSS sources; they should not be interpreted as a measurement of the intrinsic absorption of the AGN spectrum that requires knowledge of the redshift.
We also note that we did not find a large change in the inferred $N_\mathrm{H}$ values over the range of count rates that our sample covers, with the average varying by $\lesssim0.1$~dex (see Figure~\ref{fig:Rate_vs_Nh}).
Nonetheless, accounting for the impact of absorption (with an effective $N_\mathrm{H}\approx10^{22}$~cm$^{-2}$) on the X-ray spectral properties of our sources in this manner does improve the accuracy of our 2--10~keV flux measurements compared to our prior assumption of Galactic absorption only and thus improves the accuracy of our number counts measurements in Section~\ref{sec: dN/dS} below.

\subsection{Differential number counts, ${dN}/{dS}$, using the hard-band ExSeSS sample}\label{sec: dN/dS}

Another way to analyse the X-ray source population is to determine the rate of change of the number counts as a function of flux, $\frac{dN}{dS}$.
This allows us to more accurately diagnose the shape of the relation for comparison with previous measurements and ensures that the errors are independent at a given flux.
As in Section~\ref{sec: Variable_column} above, here we focus on the hard (2--10~keV) selected sample only.

Following the methods of \citet[][]{mateos2008high}, the differential number density of sources per unit flux and sky area, $n(S_j)$, are given by
\begin{equation}\label{equation: 2}
n(S_{j}) = \frac{dN}{dS} =  \frac{\displaystyle\sum_{i=1}^{i=m} \dfrac{1}{\Omega_i}}{\Delta S_{j}} \end{equation}
 where $m$ is the number of sources in bin $j$ with assigned flux $S_{j}$, $\Omega_{i}$ is the sky area coverage (in deg$^{2}$) for source $i$, and $\Delta S_{j}$ is the width of the bin in flux.
The errors in $n(S_j)$ are calculated in a similar way as in Section \ref{sec: logN-logS} based on Poisson statistics, thus the error in each binned measurement is given by $n(S_{i})/m^{\frac{1}{2}}$ \citep[][]{mateos2008high}.
We note that our bins are initially defined in terms of count rate. The sky area, $\Omega_i$, is also determined depending on the count rate so that it is independent of the spectral assumption.
If a bin contains less than 10 sources it is merged with the previous bin and the average count rate of sources in the combined bin is used to calculate $S_j$; otherwise, the central point of the bin in count rate is used to calculate $S_j$.
Once the bins have been defined then count rates are converted to 2--10~keV fluxes assuming a spectrum with $\Gamma=1.9$ and the variable $N_\mathrm{H}$, depending on the hard-band count rate, determined in Section~\ref{sec: Variable_column} above.

\begin{figure}
	\includegraphics[width=\columnwidth]{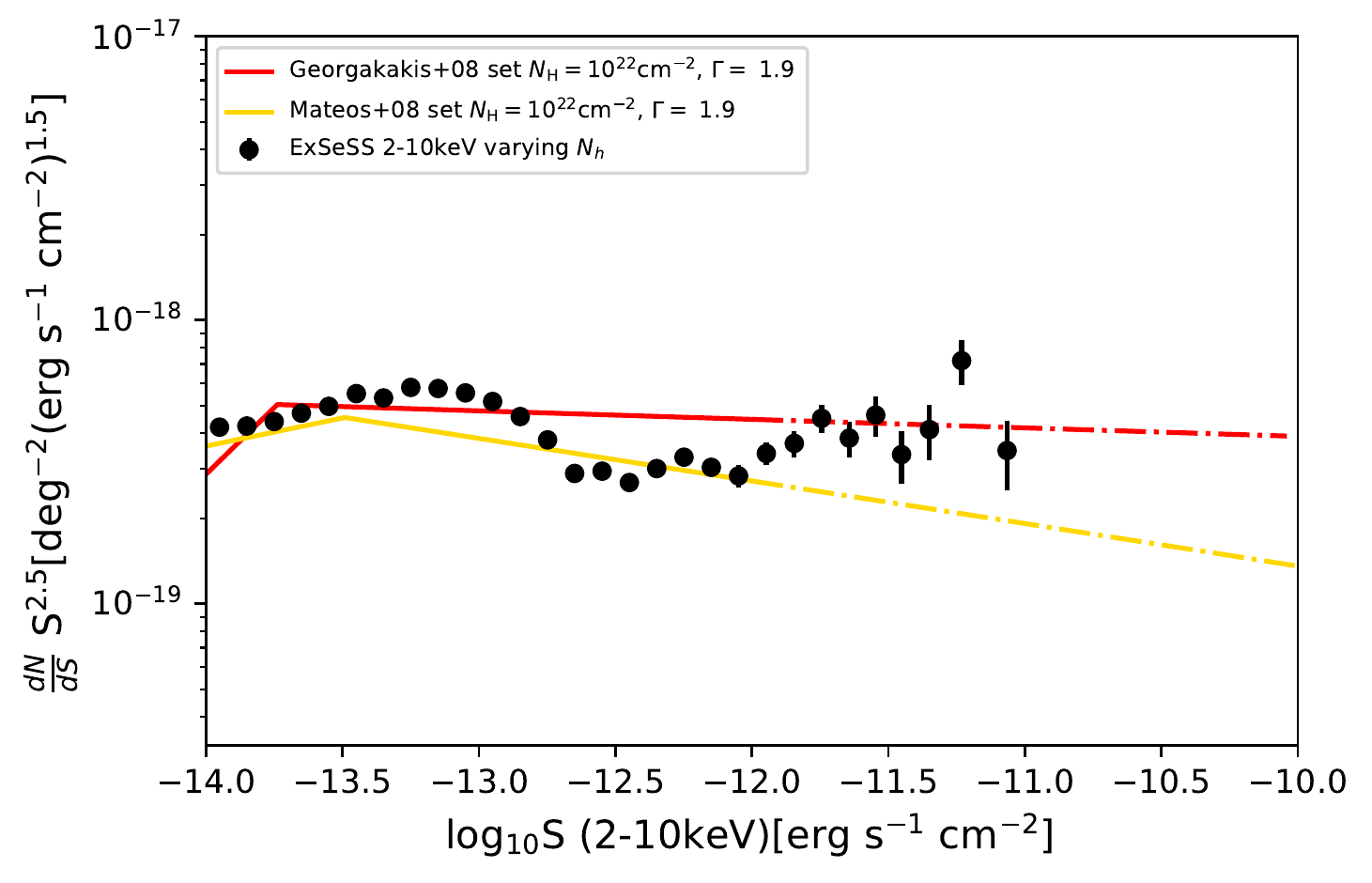}
    \caption{ExSeSS hard band (2--10~keV) sample compared with the fits from \citet[][]{mateos2008high} using \textit{XMM-Newton} data and \citet[][]{georgakakis2008new} using \textit{Chandra} data, in both cases assuming a fixed $N_\mathrm{H} = 10^{22}$ and $\Gamma = 1.9$ when estimating fluxes. The dot-dashed portions of the lines indicate where the best fitting relation has been extrapolated beyond the range of fluxes probed by a given survey. ExSeSS follows similar trends as \citet[][]{georgakakis2008new} up to $S_\mathrm{2-10keV}\sim10^{-12.75}$~ergs~s$^{-1}$~cm$^{-2}$ where ExSeSS then decreases to agree more closely with the \citet[][]{mateos2008high} fit, with good agreement at $S_\mathrm{2-10keV}\sim 10^{-12.5}-10^{-12.0}$~erg~s$^{-1}$~cm$^{-2}$. At brighter fluxes there is increased scatter in the ExSeSS measurements but they rise again to agree with the extrapolation of the \citet[][]{georgakakis2008new} fit.}
    \label{fig:dN_dS_fit_comparison}
\end{figure}

Figure \ref{fig:dN_dS_fit_comparison} shows our measurements of $\frac{dN}{dS}$ normalised relative to the Euclidean slope, i.e. $\frac{dN}{dS} \times S^{2.5}$.
Normalising by the Euclidean slope allows for even more subtle variations to be identified between different measurements of the differential number counts that have a steep slope over this flux range. We will normalise by the Euclidean slope for all our plots of $\frac{dN}{dS}$.
Figure~\ref{fig:dN_dS_fit_comparison} compares our ExSeSS results with the double power-law fits from \citet[][based on a compilation of dedicated \textit{Chandra} surveys]{georgakakis2008new} and \citet[][based on a well-defined subset of 2XMM]{mateos2008high}.
We find good agreement between ExSeSS and the \citet[][]{georgakakis2008new} relation at faint fluxes ($S_\mathrm{2-10keV}\sim10^{-14} - 10^{-12.75}$erg s$^{-1}$~cm$^{-2}$). At $S_\mathrm{2-10keV}\sim10^{-12.75} - 10^{-12.0}$erg s$^{-1}$~cm$^{-2}$ the differential number counts dip and agree more closely with the \citet[][]{mateos2008high} fit before increasing again to match the \citet[][]{georgakakis2008new} fit again at brighter fluxes (albeit with larger uncertainties in this regime).
We note that the \textit{Chandra} and \textit{XMM-Newton} studies cover substantially smaller areas than ExSeSS and thus are dominated by sources at relatively fainter fluxes. The dot-dashed lines in Figure \ref{fig:dN_dS_fit_comparison} show where the fits have been extrapolated beyond the flux range covered by the corresponding source samples.
The overall agreement between ExSeSS and these previous studies is a positive sign but the changing shape over the full flux range that we probe with ExSeSS hints that the AGN population may be more complex than previously thought. Given the high precision of ExSeSS, it is clear that a simple power law does \emph{not} describe the differential number counts well in this flux regime.

\subsection{Reconciling differential number counts with measurements at higher energies}
\label{sec:higherenergies}

\begin{figure}
    \centering
    {{\includegraphics[width=.45\textwidth]{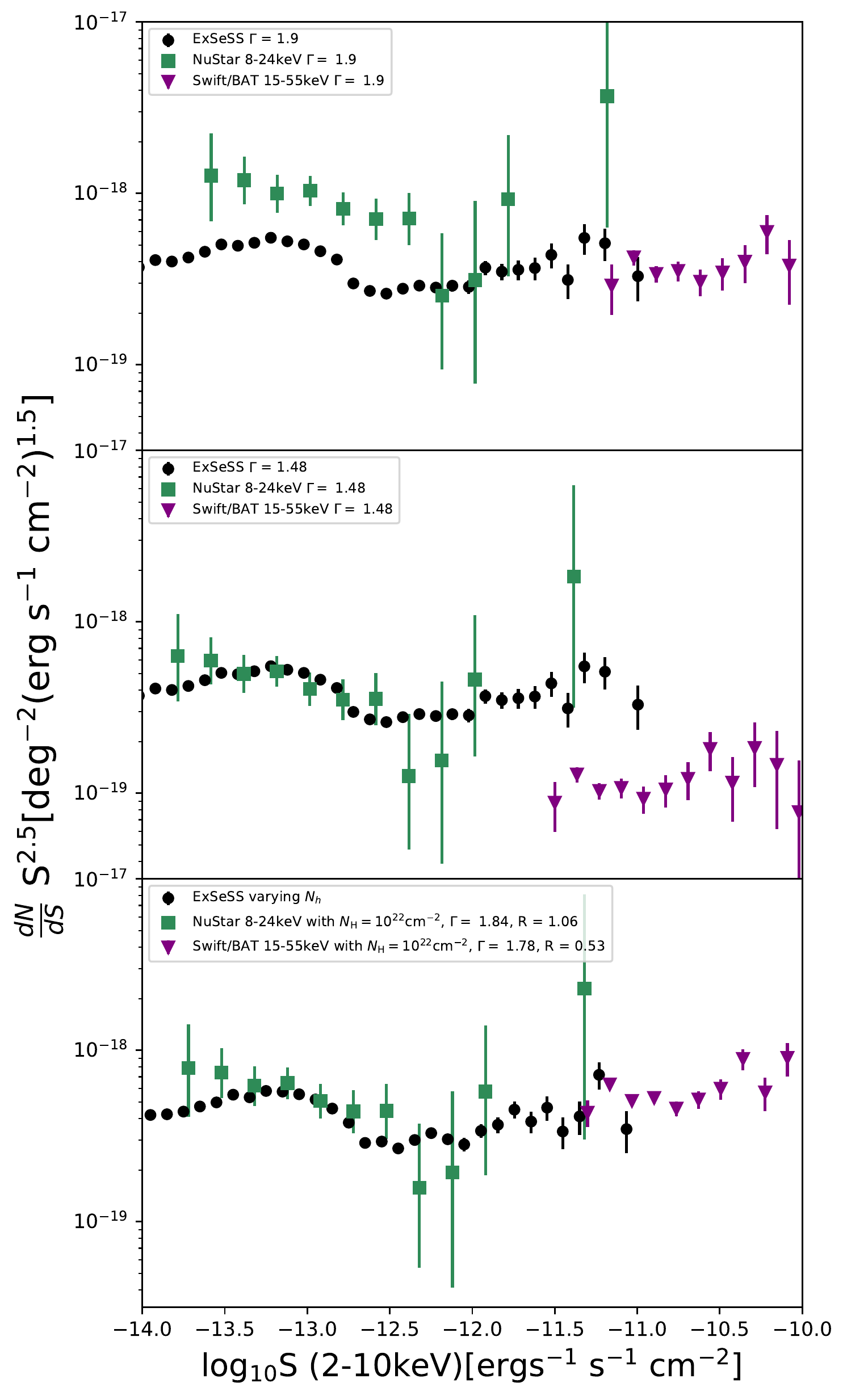} }}
    \caption{The top panel is a comparison of the ExSeSS sample to the \textit{NuSTAR} 8-24~keV and \textit{Swift}/BAT 15-55~keV surveys \citep[][]{harrison2016nustar} using a underlying spectral assumption of single power law with a  photon index of $\Gamma = 1.9$. The middle panel is a similar comparison as in the first one but using a fixed $\Gamma = 1.46$ and therefore assuming we are observing a less obscured population for \textit{NuSTAR} and \textit{Swift}/BAT. Using a variable $N_\mathrm{H}$ for ExSeSS and the correct underlying spectral properties for \textit{NuSTAR} and \textit{Swift}/BAT gives a much better agreement between the ExSeSS sample and the previous results when compared with just using a simple photon index as shown in the bottom panel. }
    \label{fig:dN_dS_data_comparison}
\end{figure}

In Figure \ref{fig:dN_dS_data_comparison}
we now compare the ExSeSS differential number counts with measurements at higher energies
from \textit{NuSTAR} \citep[at 8--24~keV:][]{harrison2016nustar} and \textit{Swift}/BAT \citep[at 15--55~keV:][]{ajello201260} and investigate the impact of different spectral assumptions that are required to translate the fluxes to the 2--10~keV band.
\refone{To calculate the equivalent 2--10~keV fluxes for each of the higher energy samples, we first translate the fluxes in the original energy bands back to count rates using the same spectral assumptions as the relevant study, and then calculate a new count-rate to 2--10~keV flux conversion factor using our own spectral assumptions (as detailed below) to estimate the equivalent flux for comparison with the ExSeSS measurements. This process thus accounts for the differing sensitivity of a given instrument over the observed energy band and the impact of the assumed spectral model on the original flux estimates.}

The top panel of Figure \ref{fig:dN_dS_data_comparison} shows results using just a simple photon index of $\Gamma$ = 1.9 (and Galactic $N_\mathrm{H}$ only) for the flux conversion.
Under this assumption, we find that the ExSeSS measurements are well aligned with the \textit{Swift}/BAT measurements at bright fluxes (although we note the limited overlap in flux range), whereas the \textit{NuSTAR} measurements at fainter fluxes are significantly higher than the ExSeSS results.
The discrepancy likely indicates a significant population of obscured AGN with X-ray spectra that are not well described by a simple $\Gamma=1.9$ power law that are identified by \textit{NuSTAR} at 8--24~keV and thus extrapolating fluxes without allowing for absorption effects leads to an over-estimate of the 2--10~keV fluxes and a discrepancy in the number counts.
In the second panel of Figure~\ref{fig:dN_dS_data_comparison} we instead assume an X-ray spectrum with $\Gamma=1.48$. \citet{harrison2016nustar} found that this flatter spectral shape, allowing for the presence of absorbed sources, was needed to bring the \textit{NuSTAR} 8--24~keV number counts into good agreement with previous, lower energy measurements with \textit{Chandra} and \textit{XMM-Newton}.
With this same spectral assumption, we find good agreement between the ExSeSS\footnote{For consistency, we also assume $\Gamma=1.48$ when calculating the ExSeSS source fluxes, although this change has a minimal impact given the count rates are measured in the same 2--10~keV band used for the flux.} and \textit{NuSTAR} measurements at fluxes $\sim10^{-14}-10^{-12}$~erg~s$^{-1}$~cm$^{-2}$, whereas at brighter fluxes there is now a significant offset between ExSeSS and the \textit{Swift/BAT} measurements.
We thus conclude that there are significant differences in the underlying X-ray spectral properties of the source samples selected by ExSeSS, \textit{NuSTAR} and \textit{Swift}/BAT at different flux and energy ranges such that a single spectral assumption is insufficient to reconcile measurements of the differential number counts.

In the bottom panel of Figure \ref{fig:dN_dS_data_comparison} we present the ExSeSS measurements using a variable $N_\mathrm{H}$ as described in Section \ref{sec: Variable_column} as well as using the \emph{observed} spectral properties of the \textit{NuSTAR} and \textit{Swift}/BAT samples when converting from the measured energy range to 2--10~keV.
For NuSTAR we adopt $\Gamma = 1.84$, $N_\mathrm{H} = 10^{22}$~cm$^{-2}$, $z = 0.58$ and include a reflection component (modeled using \textsc{pexrav}) with a relative normalization of  $R = 1.06$, based on the mean of the parameters determined from spectral analysis of the \textit{NuSTAR} 8--24~keV selected sample by \citet[][]{zappacosta2018nustar}.
For \textit{Swift}/BAT we adopt $\Gamma = 1.78$, $N_\mathrm{H} = 10^{22}$~cm$^{-2}$ and a reflection component with $R = 0.53$, based on the mean values from the spectral analysis by \citet[][]{ricci2017bat}.
Conversion factors to translate fluxes between energy bands, assuming the above parameters, were calculated using XSpec\footnote{https://heasarc.gsfc.nasa.gov/xanadu/xspec/}.
\refone{We note that our method adopts a single spectral assumption for \emph{all} sources in a given high-energy sample, while in reality there will be a distribution of properties across the population that may have second-order effects on the extrapolated 2--10~keV differential number counts. Nonetheless, using this method we obtain} a good agreement between ExSeSS, \textit{NuSTAR} and \textit{Swift}/BAT, in contrast to the results shown in the top and middle panels that assumed a single photon index showing how important it is to take into account the underlying spectral properties of the sources identified by different surveys when comparing measurements of the number counts.
Our approach means that we adopt the closest spectral properties to the population probed in a given sample, but we have not further tuned other parameters to make the different samples agree with each other.

It is interesting to note that the spectral parameters are different for \textit{NuSTAR} and \textit{Swift}/BAT with \textit{Swift}/BAT sources typically having half the reflection component of NuSTAR and the photon index differing by 0.06 hinting at differences in the AGN population identified at higher energies in different flux ranges. Applying the appropriate flux conversions brings the \textit{Swift}/BAT and \textit{NuSTAR} measurements of differential number counts into good agreement with our measurements with ExSeSS.

\subsection{Comparison with AGN population synthesis models}
\label{sec:popsynthmodels}

\begin{figure}
    \centering
    {{\includegraphics[width=.45\textwidth]{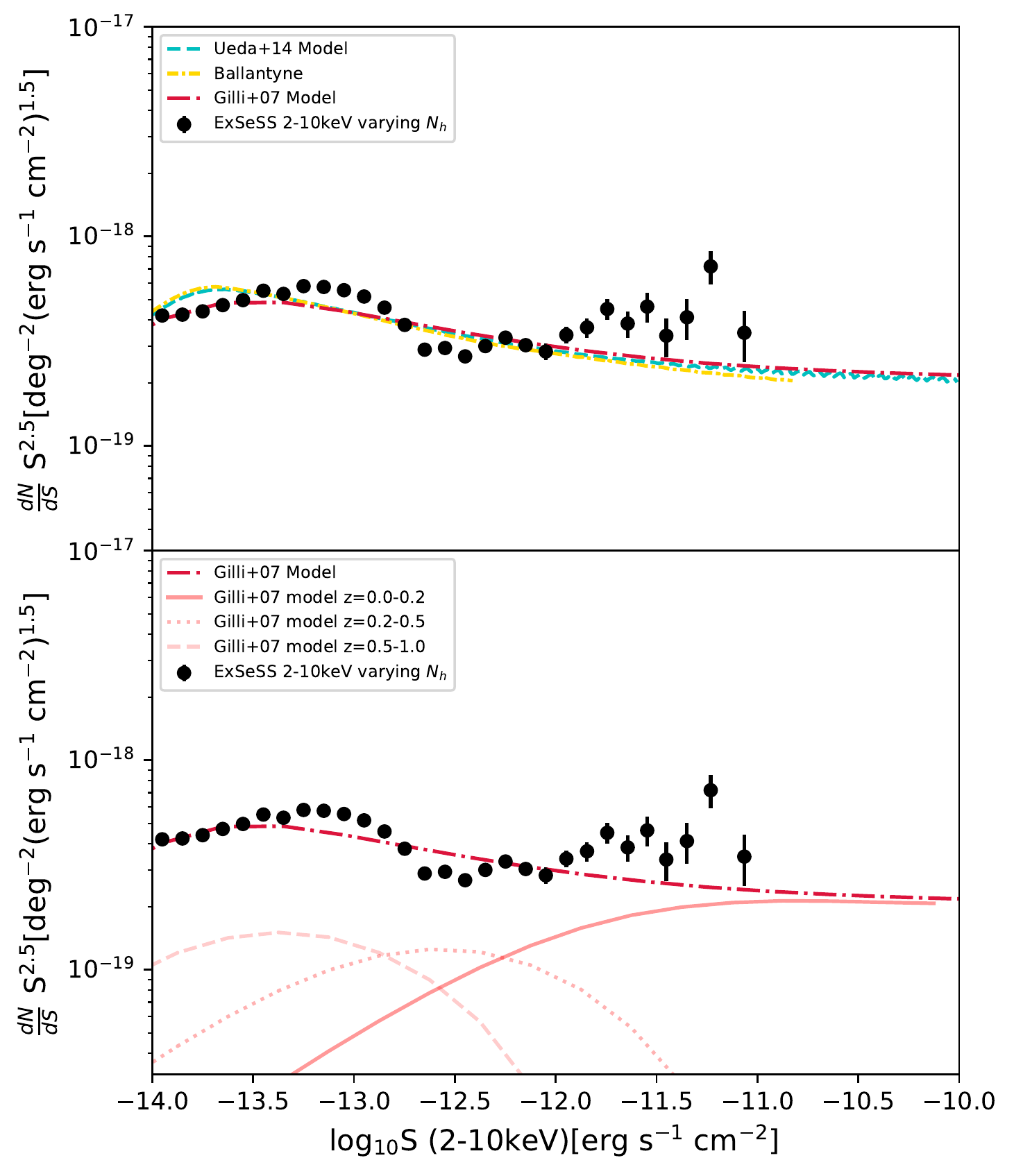} }}
    \caption{\refone{\textit{Top panel:}} Comparison of the differential number counts of the ExSeSS sample using the variable $N_\mathrm{H}$ method from Section \ref{sec: Variable_column} and the population synthesis models from \citet[][]{ueda2014toward, gilli2007synthesis} and \citet[][as updated in \citealt{harrison2016nustar}]{ballantyne2014average}. There is a generally good agreement between the ExSeSS sample and the different models \citep[][]{ueda2014toward, gilli2007synthesis} and Ballantyne over the flux range $10^{-14} - 10^{-12}$~erg s$^{-1}$ cm$^{-2}$ and between ExSeSS and \citet[][]{gilli2007synthesis} for values $<10^{-13.5}$~erg s$^{-1}$ cm$^{-2}$ . At the bright end the models and ExSeSS start to diverge at values $>10^{-12}$~erg s$^{-1}$ cm$^{-2}$. \refone{\textit{Bottom panel:} comparison of the ExSeSS measurements and contributions to the \citet[][]{gilli2007synthesis} model from AGN over different redshift ranges (solid, dotted and dashed pink lines corresponding to the indicated redshifts). At bright fluxes $>10^{-12}$~erg s$^{-1}$ the number counts are dominated by low redshift sources with $z=0.0-0.2$. This is where our main discrepancy lies and ExSeSS hints at a population of AGN at these lower redshifts that is not accounted for in the current models. }
 }    \label{fig:dN_dS_models}\end{figure}

Figure \ref{fig:dN_dS_models} shows how the differential number counts of ExSeSS and \textit{Swift}/BAT sample compare with the predictions of population synthesis models from \citet[][]{ueda2014toward, gilli2007synthesis} and \citet[][as updated in \citealt{harrison2016nustar}]{ballantyne2014average}.
As in Sections~\ref{sec: dN/dS} and \ref{sec:higherenergies} above, we adopt the variable (count-rate dependent) estimates of $N_\mathrm{H}$ when determining the ExSeSS fluxes as detailed in Section \ref{sec: Variable_column}.
\refone{The population synthesis models accout for the diverse range of spectral properties (i.e. a range of absorption columns, redshifts and luminosities) in the underlying AGN population when predicting the number counts as a function of 2--10~keV flux, enabling a direct comparison to our ExSeSS measurements.}

There is good agreement between ExSeSS and the \citet[][]{gilli2007synthesis} model at the faint end of our flux range ($\lesssim 10^{-13.5}$~erg~s$^{-1}$~cm$^{-2}$) and over the flux range $10^{-14} - 10^{-12}$~erg~s$^{-1}$~cm$^{-2}$ the ExSeSS measurements are broadly consistent with the model predictions, although the differential number counts of ExSeSS vary much more over this flux range, similar to as seen in Figure \ref{fig:dN_dS_fit_comparison} comparing with \citet[][]{mateos2008high, georgakakis2008new} power-law fits.
There are thus hints from ExSeSS that there is more variation in the differential number counts of the AGN population over this flux range than the models are currently predicting.

At high fluxes ($\gtrsim 10^{-12}$~erg~s$^{-1}$~cm$^{-2}$), the measurements indicate that there are substantially more AGN than the models currently predict meaning they need to be updated over this flux regime. This discrepancy could be due to the fact that at 2--10~keV energies at these fluxes we are detecting more AGN than predicted by the models, which are they are primarily constrained by surveys probing fainter flux regimes and lower energies. \refone{To investigate further, in the lower panel of Figure~\ref{fig:dN_dS_models} we show the contributions to the \citet[][]{gilli2007synthesis} model from AGN in different redshift ranges. It is clear that at bright fluxes ($\gtrsim 10^{-12}$~erg~s$^{-1}$~cm$^{-2}$) the model is dominated by low redshift sources, $z=0.0-0.2$.  It is thus likely that ExSeSS is picking up a population of AGN at these redshifts which the models do not currently take into account.}
\citet[][]{harrison2016nustar} also showed that there is a discrepancy between the \textit{NuSTAR} 8--24~keV results and the \textit{Swift}/BAT 15--55~keV results (converted to the 8--24~keV band) as well as showing a discrepancy between the data and the models
\refone{and suggested that updates are required to the population synthesis models at lower
redshifts.}
However, \citet[][]{harrison2016nustar} found that the models \emph{over}-predict the \textit{Swift}/BAT number counts, whereas here we find that the models are under-predicting compared to our ExSeSS measurements at 2--10~keV.
Overall our results indicate that the population synthesis models may need updating to accurately account for this population of AGN indentified at 2--10~keV energies over this flux range.

\refone{An alternative explanation for the increase in the ExSeSS number counts at bright fluxes compared to the AGN population synthesis models could be due to the presence of a contaminating, non-AGN population of sources. Without multiwavelength associations or classifications (deferred to the next stage of our study), we are unable to classify individual X-ray sources directly.
However, by extrapolating the number counts of different populations (AGN, galaxies, and stars) measured in the CDFS by \citet{lehmer2012}, we can estimate the expected level of contamination.
At a 2--10~keV flux of $S = 10^{-12.5}$~erg~s$^{-1}$~cm$^{-2}$
(the lowest point of the ExSeSS sample in Figure \ref{fig:dN_dS_models}) we measure a value of
$\frac{dN}{dS} S^{2.5} \approx 3\times 10^{-19}$~deg$^{-2}$~(erg~s$^{-1}$~cm$^{-2})^{1.5}$. Extrapolating the power-law fits from \citet[][]{lehmer2012} for the 2--8~keV band to the same flux predicts a value of $\approx 7.4 \times 10^{-21}$ for stars, $\approx 2.1 \times 10^{-21}$ for galaxies and $\approx 2.6 \times 10^{-19}$ for AGN. The number counts values we get for our sample agree well with the \citet{lehmer2012} numbers counts for AGN whereas the galaxy and star number counts are significantly lower suggesting these populations do not substantially contaminate our sample and gives us great confidence that the ExSeSS sample is dominated by AGN. It also suggests that the up-turn in our number counts measurements at bright fluxes is not caused by any star or galaxy contribution. }

\section{Summary and Conclusions}\label{sec: Conclusion}

In this paper we presented the new X-ray source catalogue for the ExSeSS sample detected in the medium (1--2~keV), hard (2-10~keV) and total (0.3--10~keV) energy bands. We use this sample to calculate the number density of sources as a function of flux, providing new constraints on the AGN population. The main points to take away from this paper are:

\begin{itemize}

    \item Based on the most recent compilation of point sources detected by the \textit{Swift}/XRT between $1^{st}$ January 2005 - $1^{st}$ August 2018 \citep[2SXPS:][]{evans20202sxps}, we have defined the Extragalactic Serendipitous Swift Survey (ExSeSS) sample with a total of 79,342 unique X-ray sources. Our sample includes sources detected in the 1--2~keV (medium), 2--10~keV (hard) and 0.3--10~keV (total) energy ranges and has a total sky coverage of $\sim$2086.6 deg$^{2}$ over the flux range of $\sim10^{-14} - 10^{-11}$erg s$^{-1}$ cm$^{-2}$. This is a previously unexplored parameter space where the ExSeSS sample can place unprecedented constraints on the X-ray source population.

    \item We compared the number counts based on the ExSeSS sample as a function of flux in the hard (2--10~keV) and total (0.3--10~keV) energy bands as well the 0.5--2~keV band (derived from the medium 1--2~keV detections) to previous results and fits shown in Figure \ref{fig:logN-logS}. There is a generally good agreement between ExSeSS and previous results, although we measure consistently higher number counts than prior studies in the medium band. We generally found good agreement with previous estimates in the hard band, although our measurements are slightly higher than \citet{masini2020chandra} and \citet{brunner2021erosita}.
    In the total band, our measurements provide accurate measurements over a wide range in flux and are in good agreement with previous measurements.

    \item
    We also compared measurements of differential number counts ($dN/dS$) from ExSeSS to
    previous
    fitted relations in the 2--10~keV band derived from \textit{Chandra} \citep[][]{georgakakis2008new} and \textit{XMM-Newton} \citep[][]{mateos2008high} (see Figure \ref{fig:dN_dS_fit_comparison}).
    Our measurements are broadly in agreement with these prior fits between $10^{-14} - 10^{-12}$~erg~s$^{-1}$~cm$^{-2}$ although our measurements show the differential number counts are not well described by a single power law over this flux range.
    Furthermore, at fluxes >$10^{-12}$~erg~s$^{-1}$~cm$^{-2}$ (where ExSeSS covers an unprecendented area compared to prior \textit{Chandra} and \textit{XMM-Newton} surveys) we see a rise in the number counts in comparison to an extrapolation of the \citet[][]{mateos2008high} fit, again showing that a power law distribution does not adequately describe the AGN number counts over this broad range of fluxes.

    \item We found a good agreement between our ExSeSS measurements of 2--10~keV differential number counts and extrapolations from previous higher energy surveys (\textit{NuSTAR} and \textit{Swift}/BAT) provided that we adopt appropriate and realistic spectral models for each sample when converting to 2--10~keV fluxes. We also require different spectral models to reconcile both the \textit{NuSTAR} 8--24~keV measurements at lower fluxes and the \textit{Swift}/BAT 15--55~keV measurements at higher fluxes with our ExSeSS 2--10~keV measurements,
    indicating there are differences in the underlying AGN populations probed by these different surveys.

    \item Comparing the hard band (2--10~keV) differential number counts to the predictions from AGN population synthesis models, we found that the ExSeSS measurements have a different shape than the model predictions.
    In particular, we found an excess in the differential number counts at fluxes $>10^{-12}$~erg~s$^{-1}$~cm$^{-2}$ with ExSeSS
        suggesting that there is an additional
        population of sources detected at harder (>2~keV) energies, \refone{likely predominantly at lower redshifts ($z\lesssim0.2$),} contributing to the number counts at bright fluxes that is not fully accounted for in current population synthesis models. \refone{This contribution corresponds to nearly twice the number density compared to that predicted by the AGN population synthesis models at the brightest fluxes covered by ExSeSS.} Hence these models may require updating over this flux range.

\end{itemize}

ExSeSS provides a new serendipitous sample of X-ray sources selected at the 0.3--10~keV, 1--2~keV and 2--10~keV energies with a total of 79,342 unique sources and an area coverage of 2086.6~ deg$^{2}$. We have used this sample to provide new constraints on the logN-logS of X-ray sources and compared these to results from other surveys performed by different telescopes, as well as comparing to the predictions of AGN population synthesis models. Comparing to other surveys we can see how the ExSeSS sample with its hard 2--10~keV sensitivity and large sky coverage over the flux range of $\sim10^{-14} - 10^{-11}$~erg~s$^{-1}$~cm$^{2}$, as shown in Figure \ref{fig:area_comparison}, can provide new constraints on the AGN population.

\refone{
The definition of ExSeSS in this paper represents an important first step toward future studies of this new, well-defined sample of X-ray sources.
In a future study, we will perform a statistical cross-match to identify counterparts to our X-ray sources at mid-infrared wavelengths \citep[using data from WISE:][]{WISE_2014} and optical wavelengths \citep[using the Legacy Survey 8:][]{Duncan_2022}. This will provide reliable multiwavelength counterparts, enabling photometric redshift estimates, and allow us to study physical properties such as the spectral properties of the AGN, obscuration properties, and host galaxy properties for a large fraction of this sample.
An SDSS-V \citep{Kollmeier2017} open-fibre programme is also underway, providing spectroscopic follow-up of the brightest sources in our hard-band sample.
Finally, we highlight work by \citet{barlow_hall_2022} using one of the sources in our ExSeSS sample that coincides with a previously known, spectroscopically identified AGN at $z=6.31$, combined with our well-defined survey sensitivity and coverage, to place constraints on the X-ray luminosity function of AGN at $z=5.7-6.4$ and demonstrating the power and utility of ExSeSS.}

\section*{Acknowledgements}

We thank the referee for helpful comments that improved this paper.
This work made use of data supplied by the UK Swift Science Data Centre at the University of Leicester.
JD and CBH acknowledge support from STFC studentships.
JA acknowledges support from a UKRI Future Leaders Fellowship (grant code: MR/T020989/1).
PAE acknowledges UKSA support.
For the purpose of open access, the authors have applied a Creative Commons Attribution (CC BY) licence to any Author Accepted Manuscript version arising from this submission.

\section*{Data Availability}

ExSeSS is based on the 2SXPS catalogue \citep{evans20202sxps}, available at \url{https://www.swift.ac.uk/2SXPS/} which provides full access to the underlying X-ray data products.
The refined sample presented here is made available as an online catalogue with this paper and through the 2SXPS website.

\appendix

\section{Source Catalogue}\label{apend:source_catalogue}

Table \ref{tab:Final_source_table} describes the final source catalogue of the ExSeSS sample with all the relevant columns. A value of $-999.0$ is used where there is no information on the source in that particular band.

\begin{table*}
\begin{center}
\begin{tabular}{ |p{3cm}||p{11cm}||p{2cm}|}
 \hline
 \multicolumn{2}{|c|}{ExSeSS column descriptions} \\
 \hline
 Column Name   & Description & Units\\
 \hline
 ExSeSS\_ID   & The ID of unique sources in the ExSeSS sample   \\
 2SXPS\_ID   & Source ID from the 2SXPS catalogue  \\
 ObsID   &  Dataset/field ID from the original 2SXPS catalogue  \\

 RA  &  The right ascension of the source position & degrees  \\
 RA\_pos  &  The positive error in right ascension of the source position & degrees  \\
 RA\_neg  &  The negative error in right ascension of the source position & degrees  \\

 Decl  &  The declination of the source position & degrees  \\
 Decl\_pos  &  The positive error in declination of the source position & degrees  \\
 Decl\_neg  &  The negative error in declination of the source position & degrees  \\

 Medium\_Rate & The rate for the medium 1-2~keV band & counts s$^{-1}$\\
 Medium\_Rate\_pos & The error in the rate for the medium 1-2~keV band & counts s$^{-1}$\\
 Medium\_Rate\_neg & The error in the rate for the medium 1-2~keV band & counts s$^{-1}$\\

 Hard\_Rate & The rate for the hard 2-10~keV band & counts s$^{-1}$\\
 Hard\_Rate\_pos & The error in the rate for the hard 2-10~keV band & counts s$^{-1}$\\
 Hard\_Rate\_neg & The error in the rate for the hard 2-10~keV band & counts s$^{-1}$\\

 Total\_Rate & The rate for the total 0.3-10~keV band & counts s$^{-1}$\\
 Total\_Rate\_pos & The error in the rate for the total 0.3-10~keV band & counts s$^{-1}$\\
 Total\_Rate\_neg & The error in the rate for the total 0.3-10~keV band & counts s$^{-1}$\\

 Medium\_Area & The corresponding area coverage to the source rate value in the medium band & deg$^{2}$ \\
 Hard\_Area & The corresponding area coverage to the source rate value in the hard band & deg$^{2}$ \\
 Total\_Area & The corresponding area coverage to the source rate value in the total band & deg$^{2}$ \\

 Soft\_Flux & The flux for the soft 0.5-2~keV band estimated from the medium (1-2~keV) band rate & erg s$^{-1}$ cm$^{-2}$\\
 Soft\_Flux\_pos & The error in the flux for the soft 0.5-2~keV band estimated from the medium (1-2~keV) band rate & erg s$^{-1}$ cm$^{-2}$\\
 Soft\_Flux\_neg & The error in the flux for the soft 0.5-2~keV band estimated from the medium (1-2~keV) band rate & erg s$^{-1}$ cm$^{-2}$\\

 Hard\_Flux & The flux for the hard 2-10~keV band & erg s$^{-1}$ cm$^{-2}$\\
 Hard\_Flux\_pos & The error in the flux for the hard 2-10~keV band & erg s$^{-1}$ cm$^{-2}$\\
 Hard\_Flux\_neg & The error in the flux for the hard 2-10~keV band & erg s$^{-1}$ cm$^{-2}$\\

 Total\_Flux & The flux for the total 0.3-10~keV band & erg s$^{-1}$ cm$^{-2}$\\
 Total\_Flux\_pos & The error in the flux for the total 0.3-10~keV band & erg s$^{-1}$ cm$^{-2}$\\
 Total\_Flux\_neg & The error in the flux for the total 0.3-10~keV band & erg s$^{-1}$ cm$^{-2}$\\

\refone{Effective} $N_\mathrm{H}$ & The \refone{effective} column density for the hard band sample (\refone{for sources with fluxes above the 0.1\% area covereage cut}) &  ~cm$^{-2}$  \\
 Hard\_Flux\_variable\_N\_h & The flux values in the 2--10~keV band calculated using the variable, \refone{effective} $N_\mathrm{H}$ as described in Section \ref{sec: Variable_column} with the same 0.1\% area cut off applied & erg s$^{-1}$ cm$^{-2}$ \\

 \hline
\end{tabular}
 \caption{Information on the columns in the ExSeSS sample including IDs from original 2SXPS data \citep[][]{evans20202sxps} being the 2SXPS\_ID, ObsID, RA, Decl and all the rate and their error columns. New columns unique to ExSeSS are the fluxes and their errors where we have the medium, hard and total converted with a simple power law of $\Gamma = 1.9$ and the hard band converted with a variable column density with a $\Gamma = 1.9$. We include the \refone{effective $N_\mathrm{H}$} values used to make the conversions for each flux value in the hard band. The value of $-999.0$ has been set for when we do not have information on the sources.
}\label{tab:Final_source_table}
\end{center}
\end{table*}

\bsp	\label{lastpage}
\end{document}